\def\cp-v{\mbox{${\hbox{$CP$\kern-0.6em\lower-.1ex\hbox{/}}}$}\, } 
\documentclass[aps,prstab,groupedaddress,floatfix,byrevtex,twoside,superscriptaddress,preprint]{revtex4}
\ifx\pdftexversion\undefined
  \usepackage[dvips]{graphicx}
\else
  \usepackage[pdftex]{graphicx}
\fi

\usepackage{color}
\usepackage{amsmath}
\usepackage[psamsfonts]{amssymb}
\usepackage{natbib}
\usepackage{hyperref}
\usepackage{pifont}
\graphicspath{{GRAPHICS-DIR/}}
\setkeys{Gin}{width=\linewidth}

\begin{document}

\preprint{BNL-75054-2005-JA}
\preprint{MUC-NOTES-327}

\title{A Cost-Effective Design for a Neutrino Factory}
\author{J.S.~Berg} \affiliation{Brookhaven National Laboratory, Upton, NY 11973, USA}
\author{S.A.~Bogacz}\affiliation{Jefferson Laboratory, 12000 Jefferson
Avenue, Newport News, VA 23606, USA}%
\author{S.~Caspi}\affiliation{Lawrence Berkeley National Laboratory,
Berkeley, CA 94720, USA}
\author{J.~Cobb}\affiliation{Oxford University, Oxford OX1 3RH, UK} 
\author{R.C.~Fernow} \affiliation{Brookhaven National Laboratory, Upton, NY 11973, USA}
\author{J.C.~Gallardo} \affiliation{Brookhaven National Laboratory, Upton, NY 11973, USA}
\author{S.~Kahn} \affiliation{Brookhaven National Laboratory, Upton, NY 11973, USA}
\author{H.~Kirk}\affiliation{Brookhaven National Laboratory, Upton, NY 11973, USA}
\author{D.~Neuffer} \affiliation{Fermi National Accelerator Laboratory, Batavia, IL 
60510, USA} 
\author{R.~Palmer}  \affiliation{Brookhaven National Laboratory, Upton, NY 11973, USA}
\author{K.~Paul} \affiliation{Muons Inc., Batavia, IL 60510, USA} 
\author{H.~Witte}\affiliation{Oxford University, Oxford OX1 3RH, UK} 
\author{M.~Zisman} \affiliation{Lawrence Berkeley National Laboratory,
Berkeley, CA 94720, USA}

\date{\today}
\begin{abstract}
There have been active efforts in the U.S., Europe, and Japan on the 
design of a Neutrino Factory. This type of facility produces intense beams 
of neutrinos from the decay of muons in a high energy storage ring. In the 
U.S., a second detailed Feasibility Study (FS2)~\cite{fs2} for a Neutrino Factory was 
completed in 2001. Since that report was published, new ideas in bunching, 
cooling and acceleration of muon beams have been developed. We have 
incorporated these ideas into a new facility design, which we designate as 
Study 2B (ST2B), that should lead to significant cost savings over the FS2 design.

\end{abstract}
\maketitle

\section{Introduction\label{sec1}}
A Neutrino Factory~\cite{geer98,status_report,blondel} facility  offers an 
exciting option for the long-term neutrino physics program. In the
U.S. there has been a significant 
investment in developing the concepts and technologies required for such an
 accelerator complex. New accelerator technologies offer the possibility of
 building, in the not-too-distant future, an accelerator complex to produce
 more than $10^{20}$ muons per year~\cite{status_report}.  
It has been proposed to build a Neutrino Factory by 
accelerating the muons from this intense source to 
energies of tens of GeV, injecting them into a storage ring 
having long straight sections, and exploiting the 
intense neutrino beams that are produced by muons decaying 
in the straight sections. The decays
\begin{equation}
    \mu^{-}  \to  e^{-}\nu_{\mu}\bar{\nu}_{e}\; , \qquad 
    \mu^{+}  \to  e^{+}\bar{\nu}_{\mu}\nu_{e}
    \label{mumpdk}
\end{equation}
 yield neutrinos that are directed along the line of the straight sections.
 This allows
 them to be observed at near and far detectors, and offers exciting
 possibilities  to pursue the study of neutrino oscillations and neutrino 
interactions with exquisite precision. 

A Neutrino Factory requires an intense multi-GeV proton source capable of 
producing a primary proton beam with a beam power of 1--4~MW or more on target. 
This is the same proton source required in the medium term for Neutrino 
Superbeams; hence, there is a natural evolution from Superbeam experiments 
to Neutrino Factory experiments. 

The physics case for a Neutrino Factory will depend upon results from the 
next round of planned neutrino oscillation experiments~\cite{aps-study}. If the unknown 
mixing angle $\theta_{13}$ is small, such that 
$\sin^{2}2\theta_{13} < O(10^{-2})$, 
or if there is a surprise and three-flavor mixing does not completely 
describe the observed phenomenology, then answers to some or all of the most 
important neutrino oscillation questions will require a Neutrino 
Factory. If $\sin^{2}2\theta_{13}$ is large, just below the present upper 
limit, and if there are no experimental surprises, the physics case for 
a Neutrino Factory will depend on the values of the oscillation parameters, 
the achievable sensitivity that will be demonstrated 
by the first generation of $\nu_e$ appearance experiments, 
and the nature of the second generation of basic physics questions that 
will emerge from the first round of results. 
In either case (large or small $\theta_{13}$), in about 
a decade the neutrino community may need to insert a Neutrino Factory into 
the global neutrino plan. The option to do this in the next 10~years 
will depend upon the accelerator R\&D that is done during the intervening 
period. 

In the U.S., the \textit{Neutrino Factory and Muon Collider Collaboration} 
(referred to
herein as the NFMCC~\cite{MC}) is a 
collaboration of 130 scientists and engineers 
engaged in carrying out the accelerator R\&D that is needed before a Neutrino 
Factory 
could be inserted into the global plan. Much technical progress has been 
made over the last few years, and several of the required key accelerator 
experiments 
are now approved. In addition to the U.S. effort, there are active Neutrino 
Factory R\&D 
groups in Europe~\cite{UK,CERN} and Japan~\cite{JAPAN}, and much of the R\&D is 
performed and organized 
as an international endeavor. Thus, because a Neutrino Factory is 
potentially the key facility for the long-term neutrino program, Neutrino 
Factory R\&D is an important part of the \textit{present} global neutrino
program. The key R\&D experiments are seeking funding now, and will need 
to be supported if Neutrino Factories are to be an option for the future.

In this article  we describe an updated 
Neutrino Factory design that demonstrates significant progress toward 
 performance improvements and cost reduction for this ambitious facility. The 
paper is organized as 
follows. Section~\ref{sec2} describes the Neutrino Factory design
concept.  The design of the front end of the facility is described in
Section~\ref{sec5} and the accelerator chain is described in
Section~\ref{sec5-sub2}. In Section~\ref{ring} we discuss the storage
  ring and the overall performance. Required R\&D is described in 
Section~\ref{RandD}. In Section~\ref{sec9} we discuss 
the assumptions used to make the cost estimate for a
Neutrino Factory and finally, we conclude with a summary in Section~\ref{sec7}.
  
Much of the work described in this paper was performed as part of the year-long 
Study of the Physics of Neutrinos, organized by the American Physical 
Society~\cite{aps-study}.

\section{Machine Concept\label{sec2}}
\begin{figure}[ptbh!]%
\includegraphics*[width=10cm]{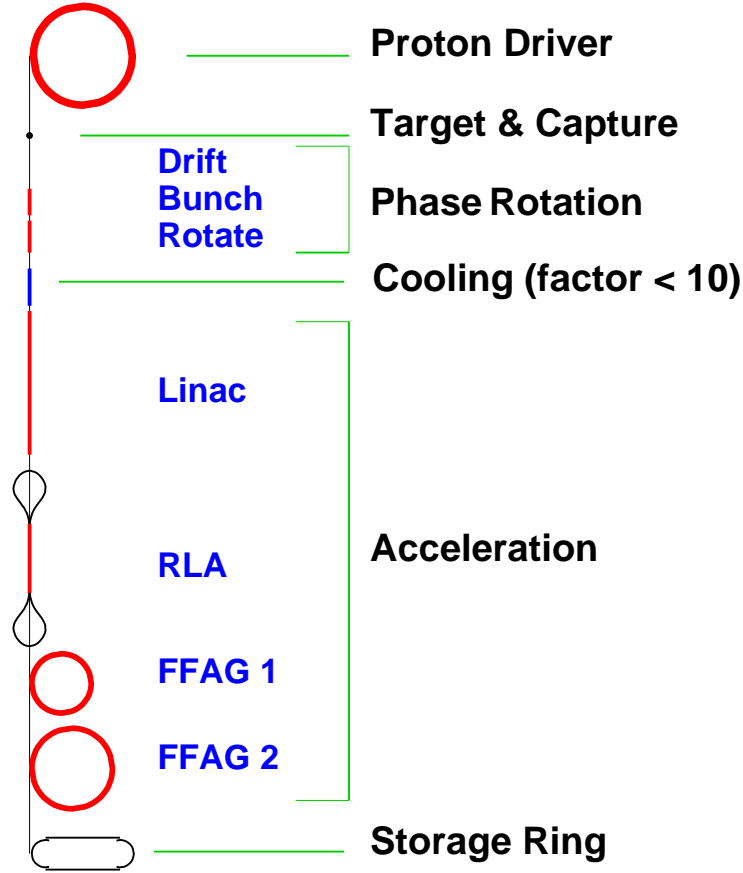}%
\caption{(Color) Schematic of a Neutrino Factory Facility.
}%
\label{schema-all}%
\end{figure}

In this Section we describe the basic machine concepts that are used to create a
Neutrino Factory facility~\cite{fs1,aps-study,fs2}; a schematic of the
whole facility is shown in Fig.~\ref{schema-all}. The facility is a \textit{quartenary beam} machine; that is, a
primary proton beam is used to create first a secondary pion beam and subsequently, a tertiary muon beam that decays
and eventually provides
the neutrino flux for the detector. For a Neutrino Factory the primary beam is a high intensity proton beam of
moderate energy (beams of 2--50 GeV have been considered by various groups)
that impinges on a target, typically a high-$Z$ material (e.g., Hg). The collisions
between the proton beam and the target nuclei produce a secondary pion beam
that quickly decays (26.0~ns) into a longer-lived (2.2~$\mu$s) muon beam. The remainder
of the Neutrino Factory is used to condition the muon beam (see Section~\ref{sec5}), accelerate it rapidly to the desired final energy of a few
tens of GeV (see Section~\ref{sec5-sub2}), and store it in a decay ring having a long straight section
oriented such that decay neutrinos produced there will hit a detector located
thousands of kilometers from the source.

Two Feasibility Studies~\cite{fs1,fs2} have demonstrated technical
feasibility (provided the challenging component specifications are met),
established a cost baseline, and established the expected range of physics
performance. 
Our present concept of a Neutrino Factory is based in part on the most recent
Feasibility Study (Study-II, referred to herein as FS2)~\cite{fs2}
that was carried out jointly by BNL and the U.S. NFMCC.  It is worth noting that the Neutrino Factory design we envision
could fit comfortably on the site of an existing laboratory, such as BNL or
FNAL. Figure~\ref{schema-all} shows a schematic of the facility. 
A summary of parameters is given in Table~\ref{tab:concept:params}.
\begin{table}[tbp]
\caption{Summary of the main parameters}
\label{tab:concept:params}%
\begin{ruledtabular}
\begin{tabular}{lc}
Proton energy (GeV) & 24\\
Driver cycle rate (Hz)&2.5\\
Bunches per spill&6\\
Average bunch rate (Hz) &$2.5 \times 6 = 15 $\\
Protons per bunch ($10^{12}$)& 1.6\\
Bunch length, rms (ns)&3\\
Proton power (MW) & 1\\
Final muon energy (GeV)&20\\
Muons of each sign per proton after cooling&0.17\\
Muons of each sign per proton after acceleration& 0.11\\
Muons of both signs per $10^7$ sec decaying toward detector&$2\times 10^{20}$\\
\end{tabular}
\end{ruledtabular}
\end{table}

The main ingredients of a Neutrino Factory include:
\begin{itemize}
\item{\textbf{Proton Driver:}} 1--4~MW (1~MW in this article, but possibly
  upgradeable to 4~MW) of protons on target from, for
  example, an
upgraded AGS; a new booster at Fermilab (or elsewhere) would perform equivalently.
\item{\textbf{Target and Capture:}} A high-power target is immersed in a 20~T
superconducting solenoidal field to capture pions produced in proton-nucleus
interactions. The high magnetic field at the target is smoothly tapered down
to a much lower value, 1.75~T, which is then maintained through the
bunching and phase rotation sections of the Neutrino Factory.
\item{\textbf{Bunching and Phase Rotation:}} We first accomplish the bunching
with rf cavities of modest gradient, whose frequencies change as we proceed
down the beam line. After bunching the beam, another set of rf cavities, with
higher gradients and again having decreasing frequencies as we proceed down
the beam line, is used to rotate the beam in longitudinal phase space to
reduce its energy spread.
\item{\textbf{Cooling:}} A solenoidal focusing channel, with high-gradient
201.25~MHz rf cavities and LiH absorbers, cools the transverse normalized rms
emittance from 17~mm$\cdot$rad to about 7~mm$\cdot$rad. This takes place at a
central muon momentum of 220~MeV/c.
\item{\textbf{Acceleration:}} A superconducting linac with solenoidal focusing
is used to raise the muon beam energy to 1.5~GeV, followed by a Recirculating
Linear Accelerator (RLA), arranged in a \textsl{dogbone} geometry, to
provide a 5~GeV muon beam. Thereafter, a pair of cascaded Fixed-Field, 
Alternating Gradient (FFAG) rings, with a triplet lattice
of combined-function magnets, is used to reach 20~GeV. Additional FFAG
stages could be added to reach a higher beam energy, if
the physics requires this.
\item{\textbf{Storage Ring:}} We employ a compact racetrack-shaped
superconducting storage ring in which $\approx35$\% of the stored muons
decay while traveling 
toward detectors located nearby, and some 3000~km from the ring. Muons survive for
roughly 500 turns.
\end{itemize}

In the remainder of this paper we describe in detail the new design of the
Neutrino Factory front-end for  performing the required beam manipulations
prior to acceleration and describe our new ideas for accelerating the muon
beam using FFAGs.

\section{Front End Design\label{sec5}}
\begin{table}[tbp]
\caption{Summary of the front-end parameters}
\label{tab:front:params}%
\begin{ruledtabular}
\begin{tabular}{lc}
Solenoid capture magnetic field (T)&20\\
Length of taper (m)&12\\
Solenoid field at end of taper (T)&1.75\\
Length of drift (m)&99\\
Solenoid field in drift (T)&1.75\\
Length of buncher (m)&50\\
Solenoid field in buncher (T)&1.75\\
Maximum rf gradient in buncher (MV/m)&10\\
Length of phase rotator (m)& 54\\
Solenoid field in phase rotator (T)&1.75\\
RF gradient in phase rotator (MV/m)&12.5\\
Length of cooler (m)&80\\
Maximum solenoid field in cooler (T)&3\\
RF gradient in cooler (MV/m)&15.25\\
\end{tabular}
\end{ruledtabular}
\end{table}
Some front end parameters are given in Table~\ref{tab:front:params}. The front 
end of the neutrino factory (the part of the facility between the
target and the first linear accelerator) represented a large fraction,
about $40\,\%,$ of the
total facility costs in FS2~\cite{fs2}. However, several recent developments
have led to  a new design for the front end
that has a crucial performance advantage and is also significantly less 
expensive. The new concepts are:
\begin{itemize}
\item A new approach to bunching and phase
rotation using the concept of adiabatic rf
bunching~\cite{adiab1,adiab2,adiab4,adiab5} eliminates the very
expensive induction linacs used in FS2.
\item For a moderate cost, the
transverse acceptance of the accelerator chain is doubled from its FS2
value. 
\item The increased accelerator acceptance diminishes the demands on the 
transverse ionization cooling and allows the design of a simplified cooling 
section with fewer components
and reduced magnetic field strength.
\end{itemize}
 We denote as \textit{Study 2B} (ST2B) the 
simulations that have been made of the performance of this new front end,
together with the new scheme for acceleration. The Monte Carlo simulations
were performed with the code ICOOL~\cite{icool}.
\begin{figure}[ptbh!]
\includegraphics*[width=12cm]{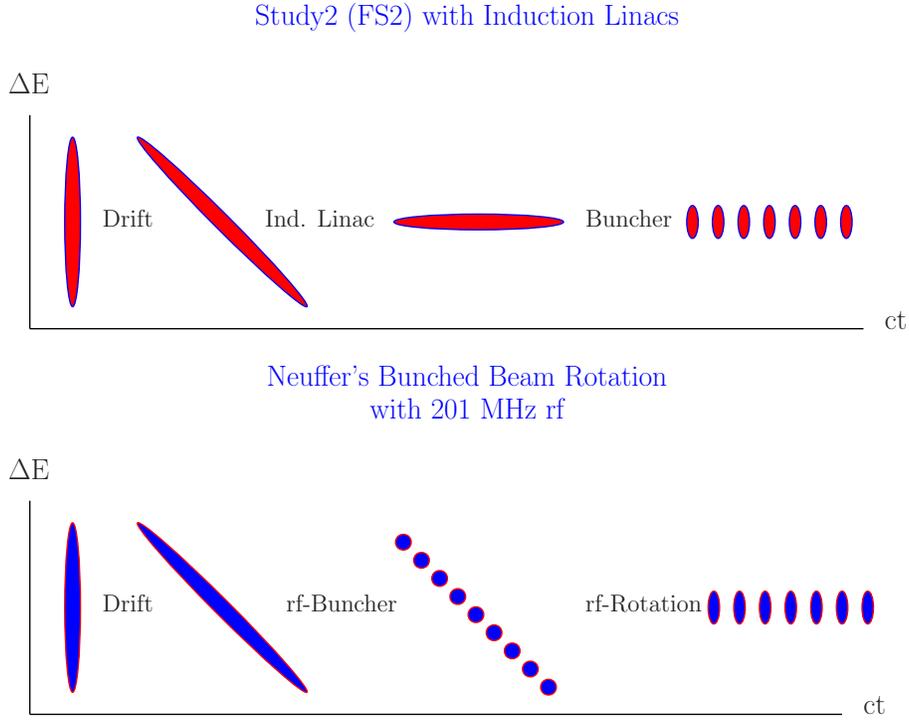}%
\caption{(Color) Comparison of the buncher concept used here (bottom) with
  the bunching system used in FS2 (top).}%
\label{fig100}%
\end{figure}
The concept of the adiabatic buncher is compared with the system used in FS2
in Fig.~\ref{fig100}. The longitudinal phase space after the target is the
same in both cases. Initially, there is a small spread in time, but a very large 
spread
in energy. The target is followed by a drift space in both cases, where a strong
correlation develops between time and energy. 
Figure~\ref{fig112new} shows the longitudinal phase space after the long
drift.
\begin{figure}[ptbh!]
\includegraphics[width=5in]{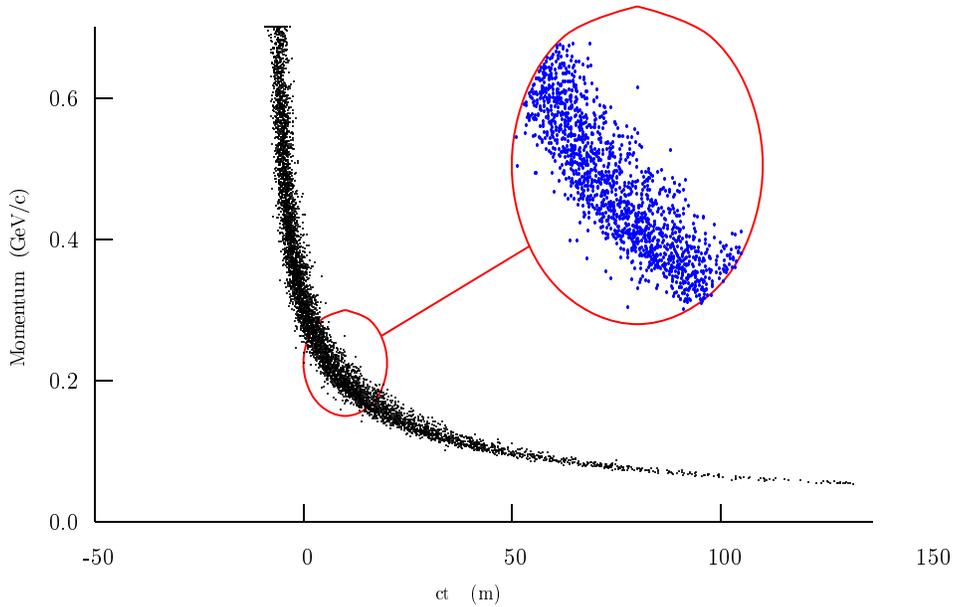}%
\caption{(Color) Longitudinal phase space after the drift section.}%
\label{fig112new}%
\end{figure}
In FS2 the energy spread in the
correlated beam was first flattened (phase rotated) using a series of induction 
linacs. The
induction linacs did an excellent job, reducing the final rms energy spread to
4.4\%, but were expensive. The beam was then sent through a series of rf 
cavities for bunching, 
which increased the energy spread to $\approx8\%.$ 

In the new scheme, the
correlated beam is first adiabatically bunched using a series of rf cavities
with decreasing frequencies and increasing gradients in such a way that the
bunch centers remain at the rf zero crossings, even as their spacing
increases because of their differing energies and velocities. 

The beam is then phase
rotated with a second string of rf cavities with decreasing frequencies and
constant gradient. In this case the frequencies are chosen with a slightly 
different criterion than that in the bunching. They are chosen 
so that the high energy bunch centers see a decelerating rf field, while
the low energy particles see an accelerating field. The final rms energy spread 
in the new design is $10.5\%.$
This spread is acceptable for the new cooling channel.
\begin{figure}[ptbh!]
\includegraphics[width=4.5in,clip]{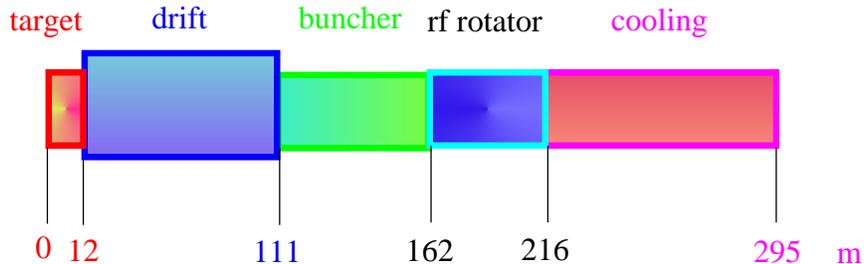}
\caption{(Color) Overall layout of the front-end.}%
\label{fig101}%
\end{figure}
The overall layout of the new front-end design is shown schematically in
Fig.~\ref{fig101}.

The first 12~m is used to capture pions produced in the target. The
field here drops adiabatically from 20~T over the target to 1.75~T. At
the same time, the radius of the beam pipe increases from 7.5~cm at
the target up to 25~cm. Next comes 99~m for the pions to decay into
muons and for the energy-time correlation to develop. The adiabatic bunching
occupies the next 50~m and the phase rotation and matching take place in 54~m
 following that. Lastly, the channel has 80~m of ionization cooling.
The total length of the new front end is $295$~m. 
\begin{figure}[ptbh!]
\includegraphics[angle=90,width=5in]{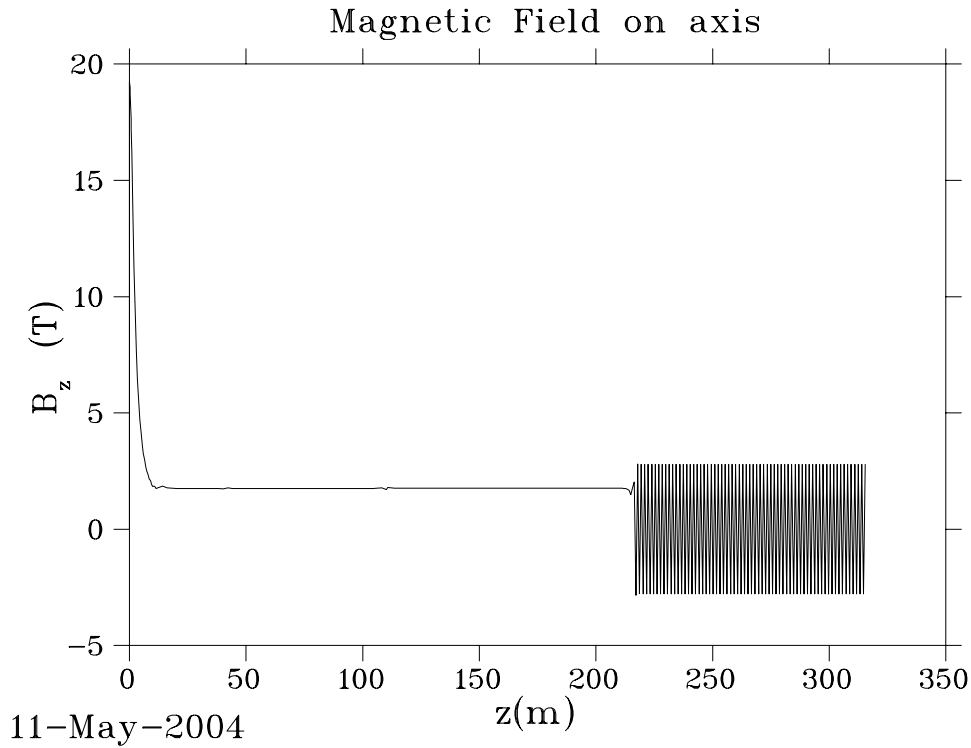}%
\caption{Longitudinal field component $B_{z}$ on-axis along the ST2B front-end.}%
\label{fig102}%
\end{figure}
The longitudinal component of the magnetic field on-axis is shown for the full 
front-end in
Fig.~\ref{fig102}. The field falls very rapidly in the collection region to a
value of 1.75~T. It keeps this value with very little ripple over the decay,
buncher and rotator regions. After a short matching section, the 1.75~T field
is changed to the alternating field used in the cooling
section.
\subsection{Target and Decay Region}
The beam distributions used in the simulations were generated using MARS~
\cite{mars1}. The distribution was calculated for a 24~GeV proton beam
interacting with a Hg jet~\cite{target}. The jet was incident at an angle of
100~mrad to the solenoid axis, whereas the beam was incident at an angle of
67~mrad to the solenoid axis. An independent study showed that the resulting
33~mrad crossing angle gives near-peak acceptance for the produced pions. An
examination of the distribution of particles that were propagated to the
end of the capture region showed that they have a peak initial longitudinal 
momentum of
$\approx 300$~MeV/c with a long high-energy tail, and a peak initial
transverse momentum $\approx 180$~MeV/c. 
\begin{figure}[ptbh!]
\includegraphics[width=4in]{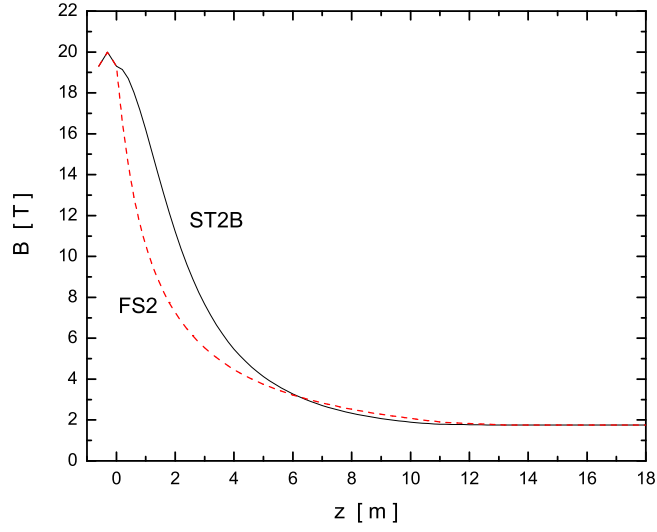}
\caption{(Color) Comparison of the
 capture region magnetic field used in the present simulation (ST2B) with
 that used in FS2.}%
\label{fig103}%
\end{figure}

\begin{figure}[ptbh!]
\includegraphics[width=4in]{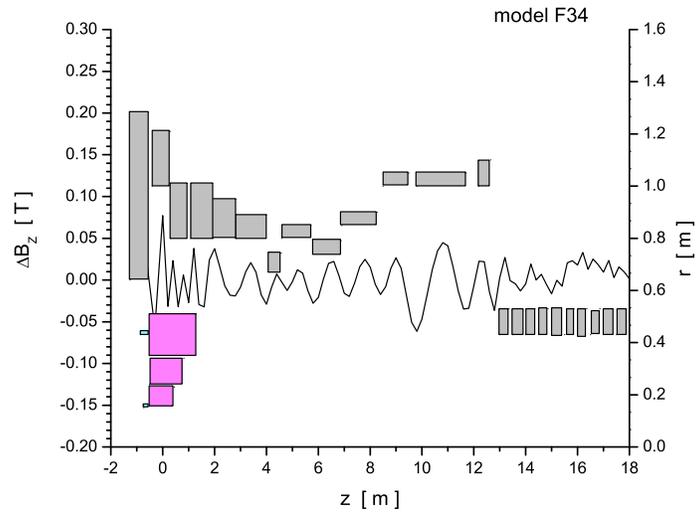}
\caption{(Color) Actual coil configuration
in the collection region. The left axis shows the error field on-axis compared
with the optimal capture field profile, denoted ST2B in Fig.~\ref{fig103}. }%
\label{fig104}%
\end{figure}

We used an improved axial field profile in the capture region that
increased the final number of muons per proton in the accelerator acceptance
by $\approx 10\%.$ The new axial field profile (marked ST2B) is compared in
Fig.~\ref{fig103} with the profile used in FS2. Figure~\ref{fig104} shows the
actual coil configuration in the collection region. The end of the 60~cm long
target region is defined as $z = 0.$ The three small-radius coils near $z=0$
are Cu coils, while the others are superconducting. The left axis shows the
error field on-axis compared with the desired field profile. We see that the
maximum error field is $\approx0.07$~T. 

\begin{figure}[ptbh!]
\includegraphics[scale=2]{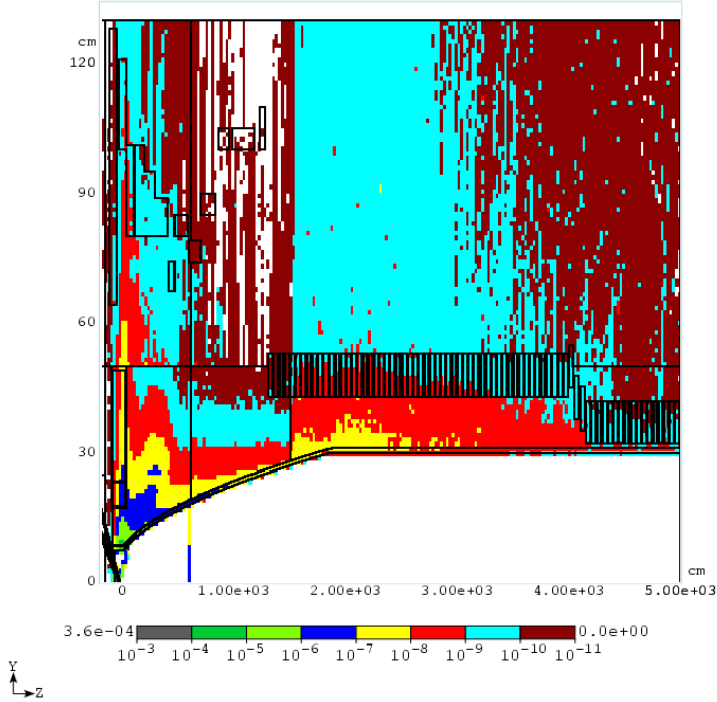}
\caption{(Color) MARS calculation
of the absorbed annual radiation dose in the collection region for a 1~MW
proton beam. The scale
shown is in units of GeV/g per proton on target.}%
\label{fig105}%
\end{figure}

Figure~\ref{fig105} shows a MARS calculation of the absorbed radiation dose in
the collection region. The peak energy deposition dose in the
superconducting coils, as illustrated in Fig.~\ref{fig105}, is $\approx
0.5\times 10^{-8}$~GeV/g per proton on target. This dose is 
$\approx1$~MGy/yr for a 1~MW beam running for a Snowmass year of
$1\times 10^{7}$~s. Assuming a lifetime dose for the insulation of 100~MGy,
 there should be no problem with radiation damage in the coils, even at a
 4~MW intensity level. 
\subsection{Bunching and Phase Rotation Region}
\begin{figure}[ptbh!]
\includegraphics[width=5in]{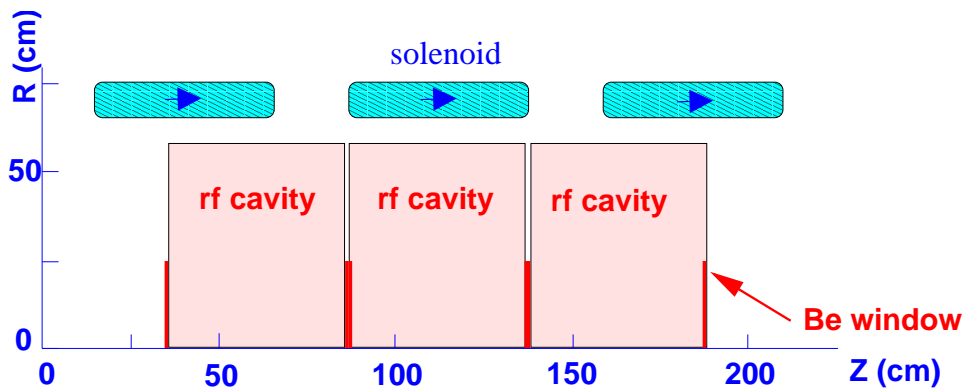}
\caption{(Color) Schematic of a cell at the beginning of the  phase rotator 
section.}%
\label{fig106}%
\end{figure}

A cell of the buncher lattice is shown schematically in Fig.~\ref{fig106}.
Most of the cell length is occupied by the 50~cm long rf cavities. The
cavity iris is covered with a Be window.  The window thicknesses varied from 
200 to 395 microns, depending on the cavity gradient. The limiting radial 
aperture in the
cell is determined by the 25~cm radius of the window. The 50~cm long solenoids
were placed outside the rf cavity with periodicity of 75~cm, in order to 
decrease the magnetic field
ripple on the axis and minimize beam losses from momentum stop bands. The
buncher section contains 27~cavities with 13~discrete frequencies and
gradients varying from 5--10~MV/m.

\begin{figure}[ptbh!]
\includegraphics[width=5in]{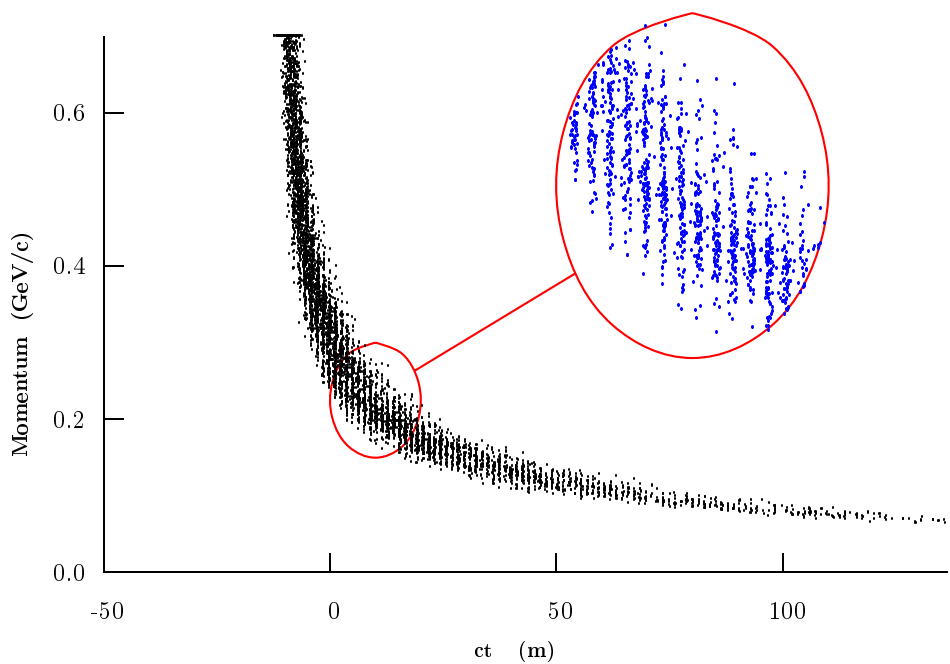}%
\caption{(Color) Longitudinal phase space of one sign muons after the buncher 
section.}%
\label{fig108}%
\end{figure}

The frequencies decrease from 333 to 234~MHz in the buncher region. The
cavities are not equally spaced. Fewer cavities are used at the beginning
where the required gradients are small. Figure~\ref{fig108} shows the
correlated longitudinal phase space and the bunching produced by the buncher.

\begin{figure}[ptbh!]
\includegraphics[width=5in]{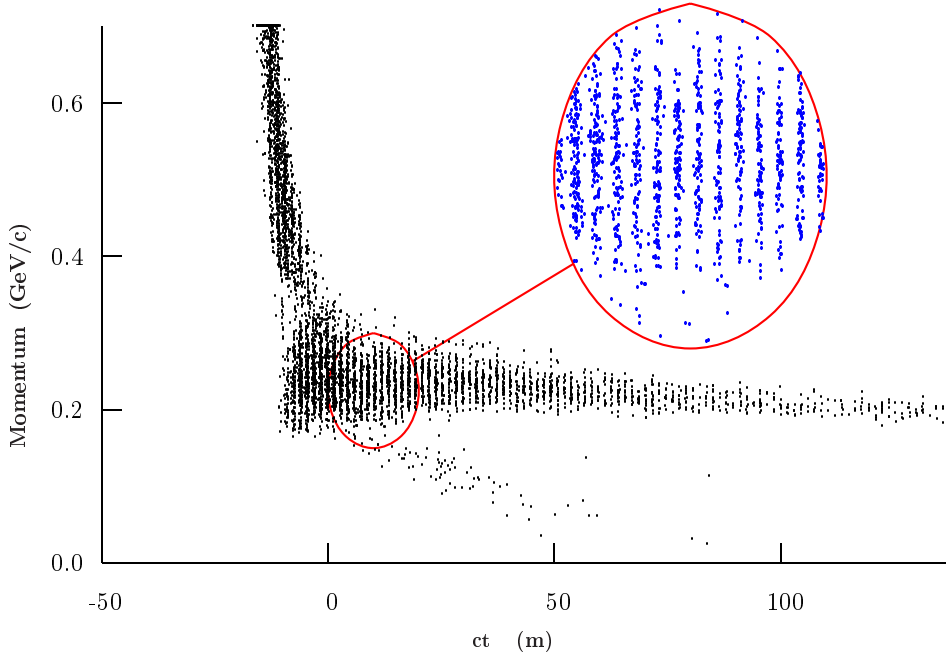}%
\caption{(Color) Longitudinal phase space after the phase rotation section.}%
\label{fig110}%
\end{figure}

The phase rotator cell is very similar to the buncher cell. The major difference 
is
the use of tapered Be windows on the cavities because of the higher rf
gradient. The tapered window had a thickness of 750~micron from the axis out 
to a radius of 20~cm and a thickness of 1.5~mm from 20 to 25~cm. There are 
72~cavities in the rotator region, with 15~different
frequencies.
The frequencies decrease from 232 to 201~MHz in this part of the front end.
All cavities have a gradient of 12.5~MV/m.
Figure~\ref{fig110} shows the longitudinal phase space after the phase rotator. 
The rms energy spread in the beam is reduced to 27~MeV.
 To study the effect of a shorter phase rotator, we also considered an example
having only a 26~m phase rotation section~\cite{NeufferRef}. This
alternative design would be
 significantly less expensive, since it is not only shorter but requires about 
200~MV less high-gradient rf voltage. Initial evaluations indicate a small
 decrease in captured muons ($\thickapprox$10\%). 
 
\subsection{Cooling Region}
\begin{figure}[ptbh!]
\includegraphics[width=4in]{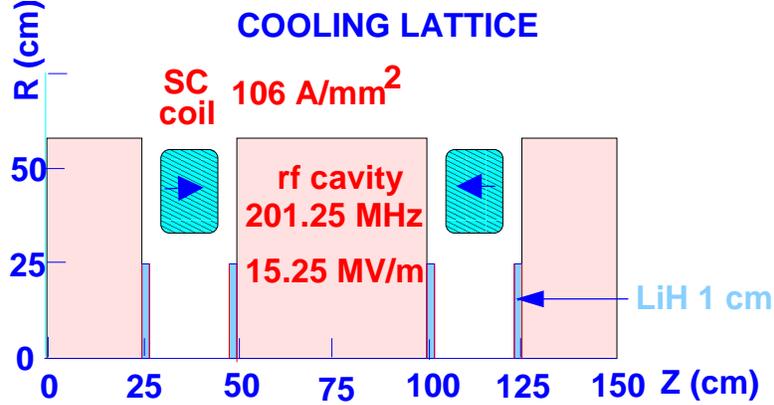}
\caption{(Color) Schematic of
one cell of the cooling section.}%
\label{fig111}%
\end{figure}

The cooling channel was designed to have a transverse beta
function that is relatively constant with position and has a magnitude of about 
80~cm. One cell of the channel is shown in
Fig.~\ref{fig111}. 
Most of the 150~cm magnetic cell length is taken up by two 50~cm long rf
cavities. The cavities have a frequency of 201.25~MHz and a gradient of
15.25~MV/m. A novel aspect of this design comes from using the windows on the
rf cavity as the cooling absorbers. This is possible because the near-constant
$\beta$ function eliminates the need to place the absorbers at a
low-$\beta$ point to prevent emittance heating. The window consists of a 1~cm 
thickness of LiH with a
$300~\mu$m thick layer of Be on the side facing the rf cavity field and a 
$25~\mu$m
thick layer of Be on the opposite side. The Be will, in turn, have a thin 
coating of TiN to prevent
multipactoring~\cite{multipac}. A 1~cm space might be introduced between
the Be rf window and LiH absorber with flowing He gas to cool both. The 
alternating 2.8~T solenoidal field is
produced with one solenoid per half cell, located between the rf
cavities. 

\begin{figure}[pbht!]
\mbox{
\includegraphics[width=0.35\linewidth,angle=90]{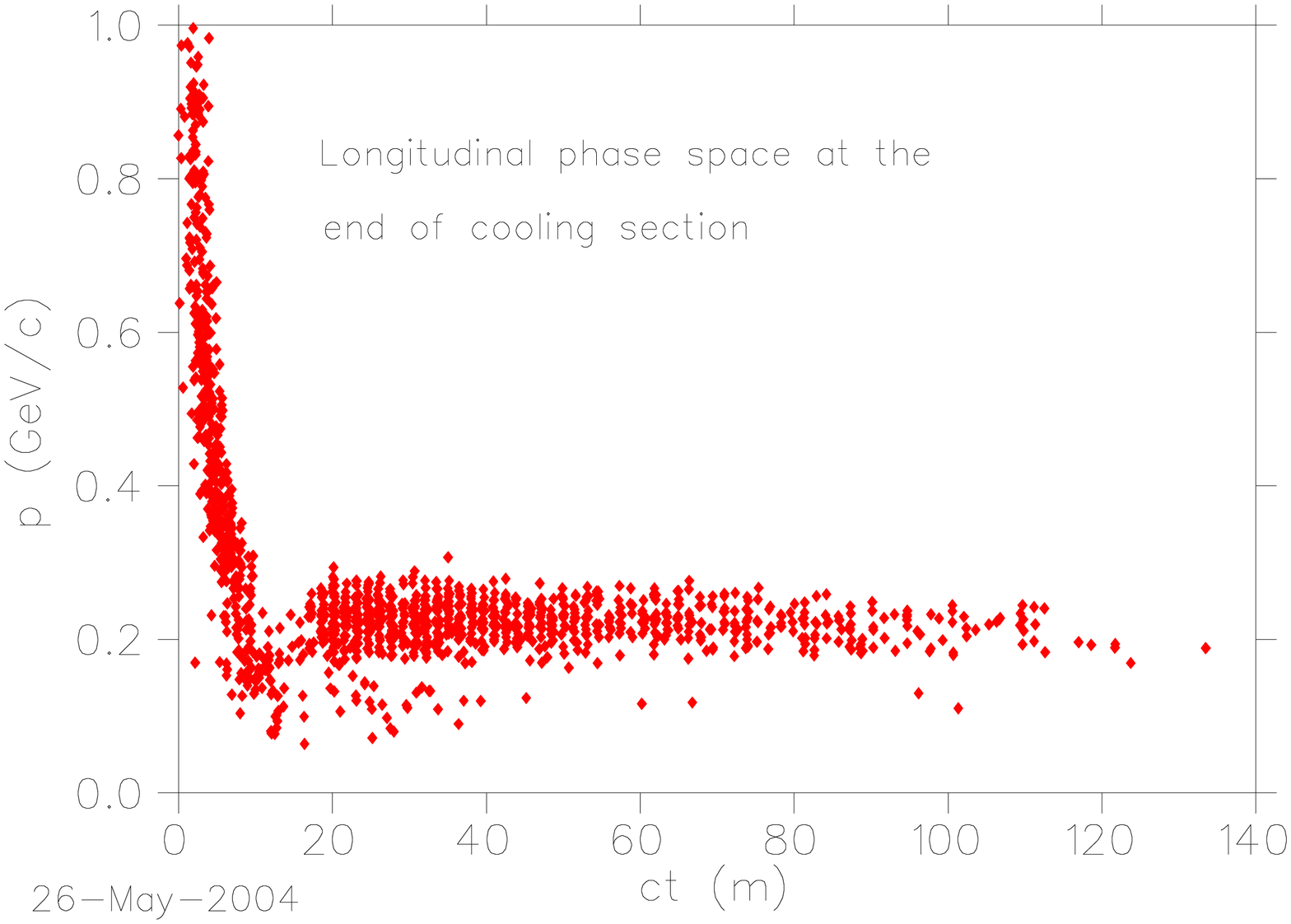}%
\includegraphics[width=0.35\linewidth,angle=90]{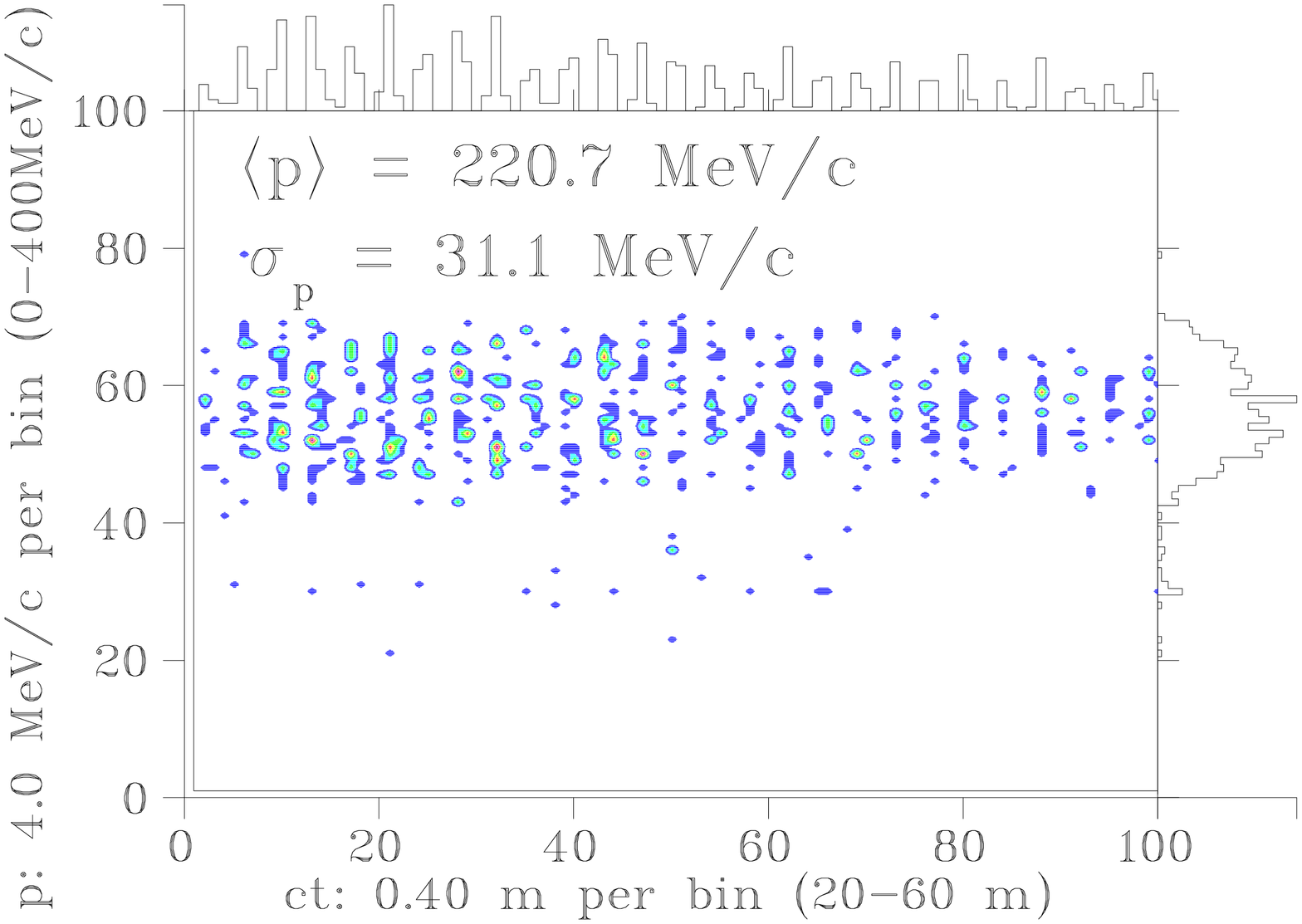}%
}\caption{(Color) Longitudinal phase space at the end of the channel.}%
\label{fig111a}%
\end{figure}

\begin{figure}[ptbh!]
\mbox{
\includegraphics[angle=90,width=3in]{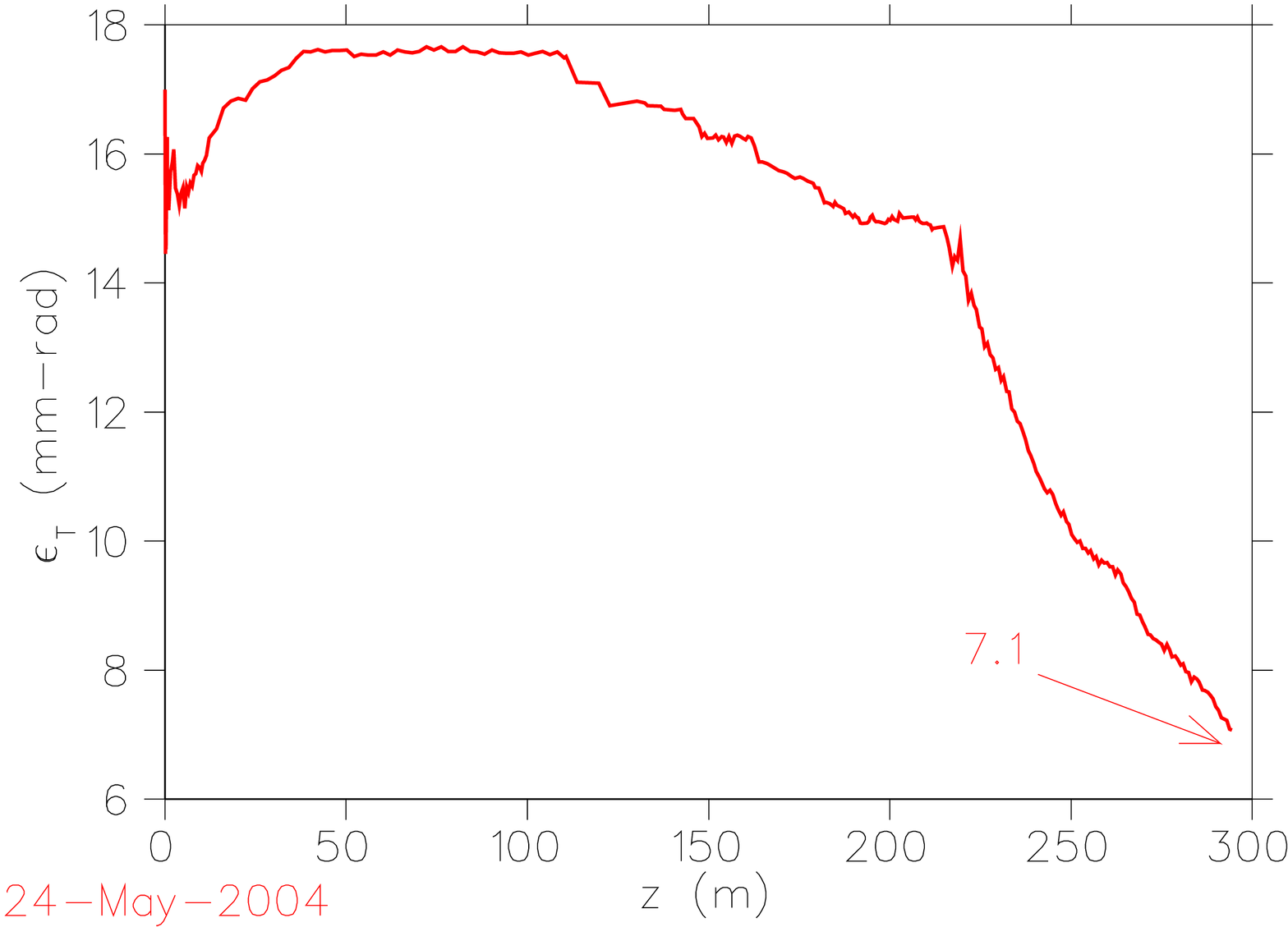}
\includegraphics[angle=90,width=3in]{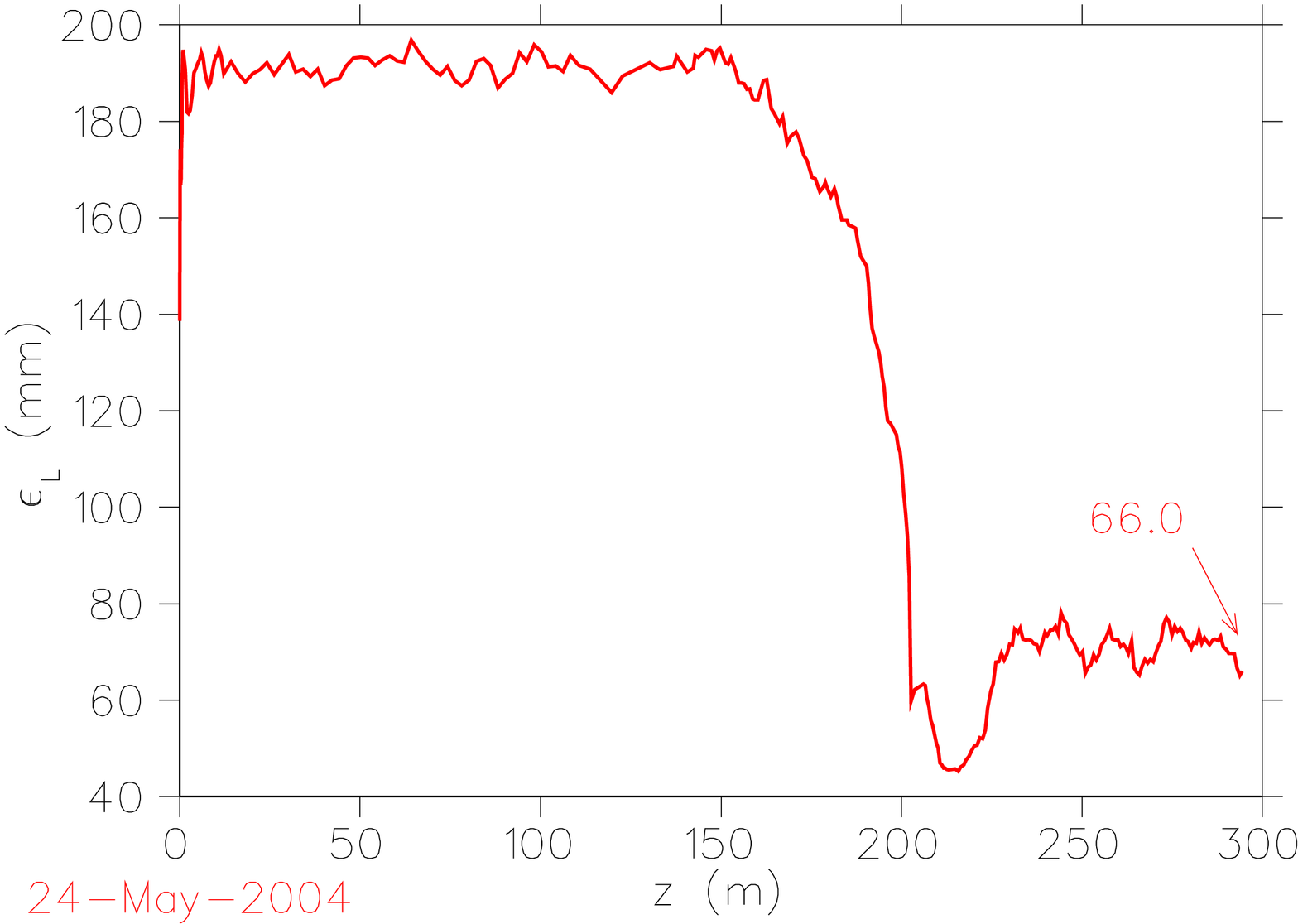}}
\caption{(Color) Normalized transverse emittance (left) and longitudinal 
emittance (right) along the front-end for a momentum cut $0.1 \leq p
\leq0.3$~GeV/c.}%
\label{fig112}%
\end{figure}

\begin{figure}[ptbh!]
\includegraphics[angle=90,width=4in]{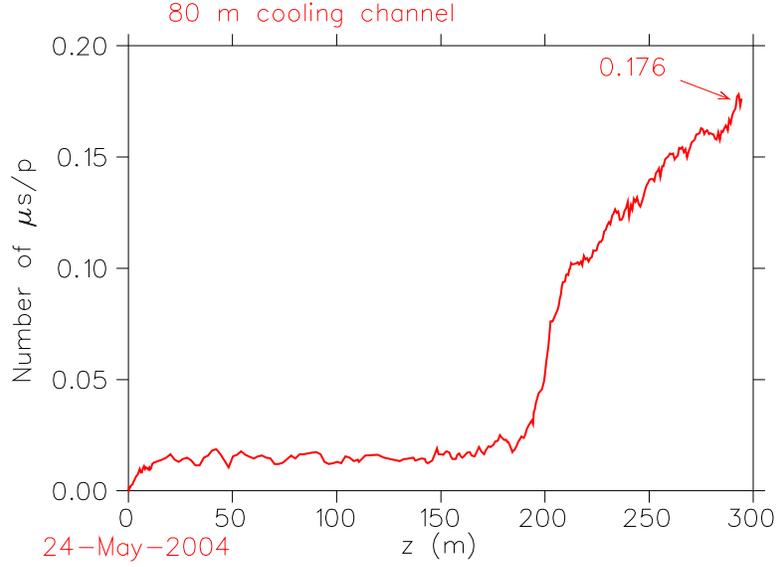}
\caption{(Color)
The muons per incident proton on target into the accelerator normalized 
transverse  acceptance of
$A_{T}=30$~mm rad and normalized longitudinal acceptance of $A_{L}=150$~mm for
a momentum cut $0.1 \leq p \leq0.3$~GeV/c.}%
\label{fig113}%
\end{figure}

\begin{figure}[tpbh!]
\vspace{0.25in}
\includegraphics[width=3.5in,clip]{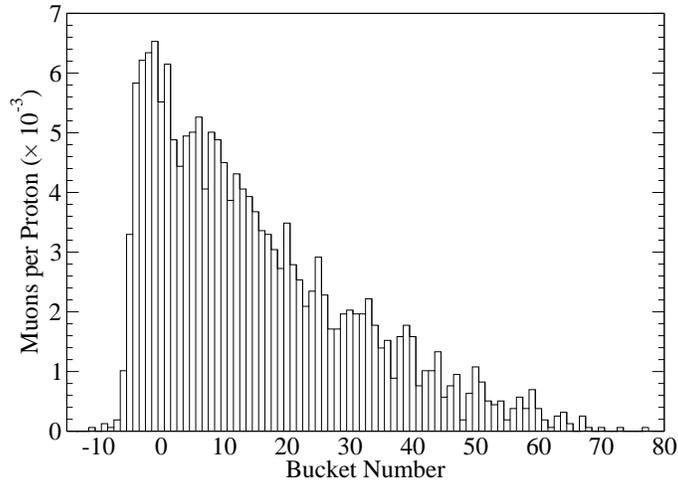}
\caption{Bunch structure
of the beam delivered to the accelerator normalized transverse  acceptance of
$A_{T}=30$~mm rad and normalized longitudinal acceptance of $A_{L}=150$~mm
for a momentum cut $0.1 \leq p \leq0.3$~GeV/c.}%
\label{fig114}%
\end{figure}

Figure~\ref{fig111a} shows the longitudinal phase space of one muon sign at the 
end
of the cooling section. The reduction in normalized transverse emittance 
$(\epsilon_T)$ along
the cooling channel is shown in the left plot of Fig.~\ref{fig112} and the
right plot shows the normalized longitudinal emittance $(\epsilon_L)$.
The channel produces a final value of $\epsilon_{T} =7.1$~mm rad, which is more 
than a
 factor of two reduction from the initial value. The equilibrium value for a LiH
absorber with an 80~cm $\beta$~function is about 5.5~mm rad. Figure~\ref{fig113} 
shows the muons per incident proton on target that fit
into the accelerator transverse normalized acceptance of $A_{T}=30$~mm rad
and normalized longitudinal acceptance of $A_{L}=150$~mm. The 80-m-long
cooling channel raises this quantity by about a factor of 1.7. The current
best value is $0.176\pm0.006$ muons per incident proton. This is the same value 
obtained in FS2. Thus, we
have achieved the identical performance at the entrance to the accelerator as
FS2, but with a significantly simpler, shorter, and presumably less expensive
channel design (see Sec.~\ref{sec9}). In addition, unlike FS2, this channel transmits both signs of
muons produced at the target. With appropriate modifications to the transport
line going into the storage ring and the storage ring itself, this design could 
deliver both (time tagged)
neutrinos and antineutrinos to the detector. 
 The beam at the end of the cooling section consists of a train of bunches  with 
a varying population of muons in each one; this is shown in
Fig.~\ref{fig114} for one sign.

\begin{figure}[ptbh!]
\includegraphics[width=4.5in]{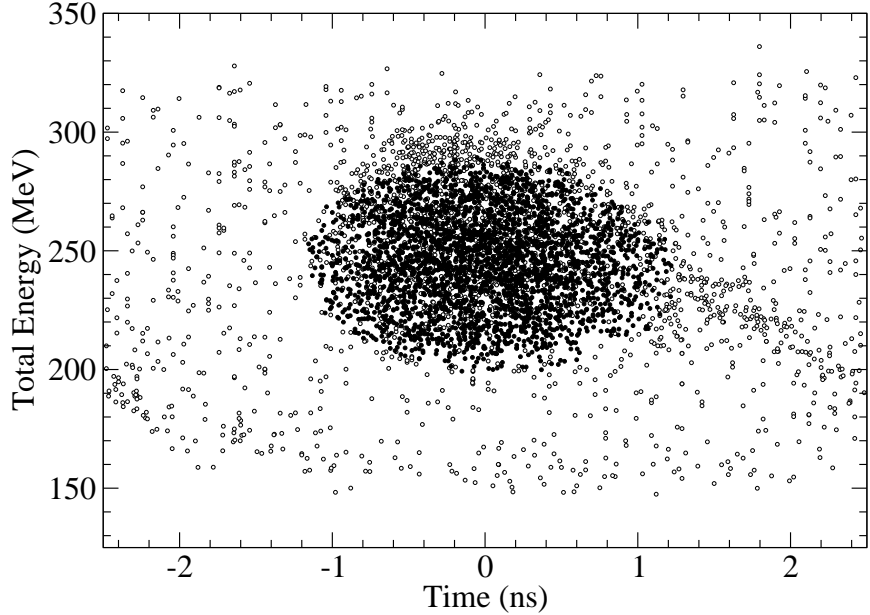}
\caption{Longitudinal phase
space of overlayed bunches in the train at the end of the cooling section. The 
open
circles are all the particles that reach the end of the channel and the filled
circles are particles within the accelerator normalized transverse 
acceptance of $A_{T}=30$~mm rad and normalized longitudinal acceptance of
$A_{L}=150$~mm for a momentum cut $0.1 \leq p \leq0.3$~GeV/c.}%
\label{fig115}%
\end{figure}

\begin{figure}[bpth!]
\includegraphics[width=4.5in]{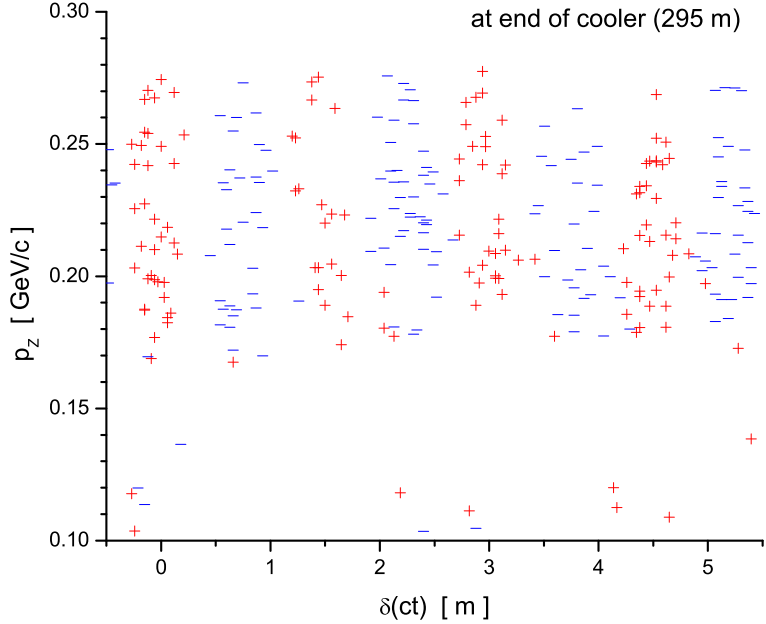}
\caption{(Color) A sample from the train of interleaved
  $\mu^{+}$~(red) and $\mu^{-}$~(blue) bunches exiting the
cooling section.}%
\label{fig116}%
\end{figure}

 Figure~\ref{fig115} depicts the longitudinal phase space of the 
 superposition of all bunches projected onto a single period $(T\approx
 5\,\text{ns})$.  We assume that particles outside the accelerator acceptance 
are intercepted 
by collimators located in the matching section, although this collimation system 
has not been designed yet. Approximately $60\%$ of the beam leaving the
cooling channel is intercepted by the collimators. The heat load on the collimator is 7.3~kW. 
Fig.~\ref{fig116} shows a few interleaved  $\mu^{+}$ and $\mu^{-}$ bunches 
exiting the cooling section. The opposite-sign bunches are mostly separated 
in time. There are a small number of wrong sign particles in the bunch train 
after cooling, but these will be cleanly separated by the dipoles in the 
subsequent accelerators and storage ring. 
\subsection{Heating of Absorber Windows}
There are some unresolved issues with the absorber windows that will require
further R\&D. To minimize multiple scattering we have assumed the windows 
are made from LiH. In order to protect the LiH from the environment and to 
provide a high conductivity surface to close off the rf cavity, we have 
assumed the LiH is encased in a thin layer of Be. Assuming that the Be can
be  bonded to the LiH, there is the question of what happens 
when the window is heated by energy loss of the muon beam and by the power 
deposited by the rf cavity. If the heating becomes high enough, melting and
 differential stresses leading to buckling are possible. In addition the window 
could suffer 
degradation from radiation damage. 

Approximately $1.1\times10^{14}$ muons of each charge enter the start of
the cooling channel each second. This produces a total power deposition of 
$\approx 58$~W distributed along the beam path. Most of the energy deposition
takes place in the LiH. 
We assume that cooling is provided by a heat sink at the outer edge of the 
window. If we ignore any longitudinal heat conduction between the LiH and
 Be layers~\cite{heating}, the LiH reaches a maximum temperature of 
$310^{\circ}$C in steady
state. This is safely below its melting temperature of  $690^{\circ}$C.  
The rf heating occurs in a skin depth on 
the side of the window facing the cavity. The skin depth for Be at 201~MHz 
is approximately $9~\mu$m.  
The rf power deposited on the window of a pillbox rf cavity is
\begin{equation}
P = \frac{\pi^2}{2} \frac{d}{\lambda } E_0^2 \frac{b^2}{Z_0}\left\{J_1^2(\alpha 
) -
  J_0(\alpha ) J_2(\alpha )\right\}
\end{equation}
where, $d=\,$skin depth, $\lambda=\,$rf wavelength, $E_0=\,$peak rf gradient,
$b=\,$window radius, $a=\,$radius of rf cavity (pillbox), $Z_0=\,$impedance
of free space, and $J_0, J_1, J_2\,$ are Bessel functions with argument
$\alpha=2.405\times\frac{b}{a}.$ This gives a total rf power of $\approx 220$~W 
in each window. Rough calculations
 predict that the temperature at the center of the $300\,\mu$m thick Be layer 
should be less 
than $175\,{}^{\circ }\text{C}.$ This is also safely below its  melting
temperature of $1275^{\circ}$C. 

Although melting will not be a problem, buckling and delamination of the Be
layer is a potential deleterious outcome. More accurate finite element
thermal studies need to be done of the composite LiH-Be system.  In case this
window design does not prove to be feasible, a
number of 
alternative absorber designs have been investigated. The window could be cooled 
by flowing He gas between the Be window and the LiH. 
The thermal conductivity to the
 heat sink on the outer edge of the window can be improved by breaking up 
the LiH into several pieces, separated by layers of high conductivity Be. 
Using a total thickness that gives the same total energy loss as the
original window results in only $\approx 3\%$ loss in the accepted muon
flux. Other possibilities that gave reasonable muon fluxes are 
a thin Be layer on pure lithium or thicker Be windows and no LiH, 
 with a
 thickness chosen to make the total energy loss the same as that in the
 baseline LiH absorber case. Cooling would be a bit less effective because of 
the greater multiple
 scattering. An initial evaluation~\cite{NeufferRef} of a
 Be-only scenario showed less capture into the acceleration channel
 acceptance ($\thickapprox$15\%). A scenario in which Be absorbers are
 initially installed and then upgraded later to more efficient LiH absorbers
 is, of course, also possible.

\section{Acceleration Design\label{sec5-sub2}}
The acceleration system takes the beam from the end of the cooling
channel and accelerates it to the energy required for the decay ring.
\begin{table}[tbp]
\caption{Acceleration system design parameters.}
\label{tab:acc:pars}%
\begin{ruledtabular}
    \begin{tabular}{lr}
Injection momentum (MeV/c) & 273\\
      Initial kinetic energy (MeV)&187\\
      Final total energy (GeV)&20 \\
      Normalized transverse acceptance (mm)&30\\
 Normalized transverse emittance, rms (mm-rad) & 3.84\\
      Normalized longitudinal acceptance, $\Delta E \Delta t/m_{\mu}c$ 
(mm)&150\\
Total energy spread, $\Delta E$ (MeV)& $\pm45.8$\\
Total time-of-flight (ns) & $\pm1.16$\\
Energy spread, rms (MeV) &19.8\\
Time-of-flight, rms (ns) &0.501 \\
      Bunching frequency (MHz)&201.25\\
      Maximum muons per bunch&$1.1\times10^{11}$\\
      Muons per bunch train (each charge)&$3.0\times10^{12}$\\
      Bunches in train&89\\
      Average repetition rate (Hz)&15\\
      Minimum time between pulses (ms)&20\\
Average beam power at the end (each charge) (kW) & 144
    \end{tabular}
  \end{ruledtabular}
\end{table}
\begin{figure}[tbp]
\includegraphics[width=\textwidth]{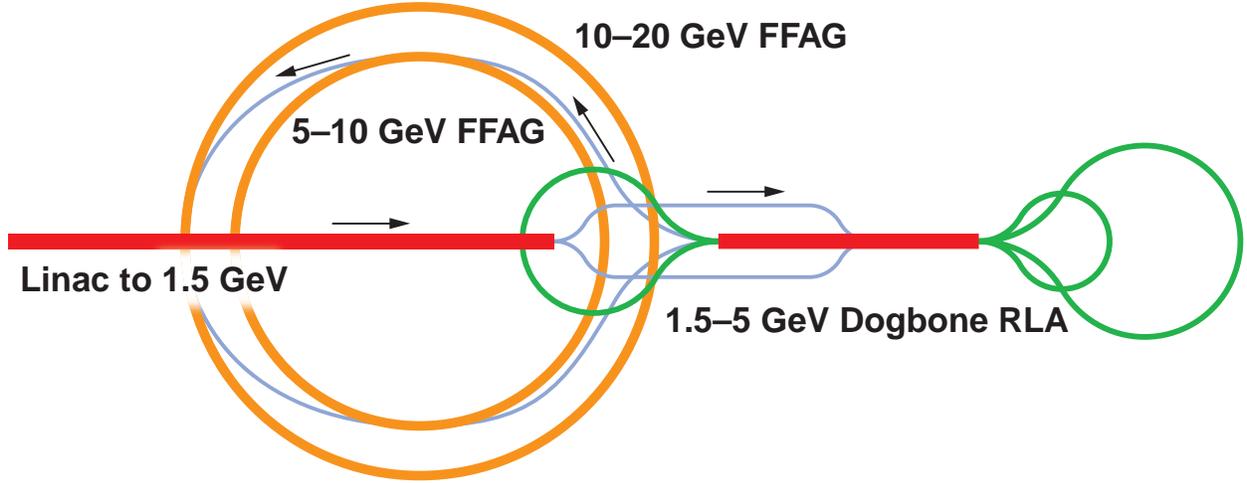}
\caption{(Color) Potential layout for the acceleration systems.} 
\label{fig:acc:alllayout}
\end{figure}
Figure~\ref{fig:acc:alllayout} shows a compact potential layout for
all the acceleration systems described here. It includes five sub-systems: a 
matching section, a linac, a Recirculating Linear Accelerator (RLA), and two 
Fixed-Field Alternating-Gradient (FFAG) circular accelerators.

To reduce costs, the RLA acceleration systems from FS2~\cite{fs2} will be
replaced, as much as possible, by Fixed-Field Alternating Gradient (FFAG) 
accelerators.
 FFAGs are rings that accelerate a beam over a large energy range
 (generally 
at least a factor of 2) without varying the magnets' fields, allowing for
very rapid acceleration.  Since they are rings, the bunches make multiple
passes through the RF cavities, reducing the RF voltage required to
accelerate.  The number of turns is not limited by the switchyard, as it is
in an RLA.  The original FFAG designs~\cite{pr103:1837} (``scaling" FFAGs)
used large, highly nonlinear magnets.  For our design, we instead use
so-called linear non-scaling FFAGs~\cite{pac99:3068,mumu4:693}.  These
FFAGs use very linear magnets to maximize the dynamic aperture (necessary
for our large-emittance beams), and the magnets generally have smaller
apertures than those in a corresponding scaling FFAG 
design, bringing down the machine cost.

Table~\ref{tab:acc:pars} gives the design parameters of the
acceleration system.  Acceptance is defined such that if $A_\bot$ is
the transverse acceptance and $\beta_\bot$ is the beta function, then
the maximum particle displacement (of the particles we transmit) from
the reference orbit is $\sqrt{\beta_\bot A_\bot mc/p}$, where $p$ is
the particle's total momentum, $m$ is the particle's rest mass, and
$c$ is the speed of light.  The acceleration system is able to
accelerate bunch trains of both signs simultaneously. The time of flight 
refers to the length of a single bunch.

\subsection{Matching from Cooling to Acceleration Linac}
\begin{figure}[!tbp]
  \centering
  \includegraphics[width=\columnwidth]{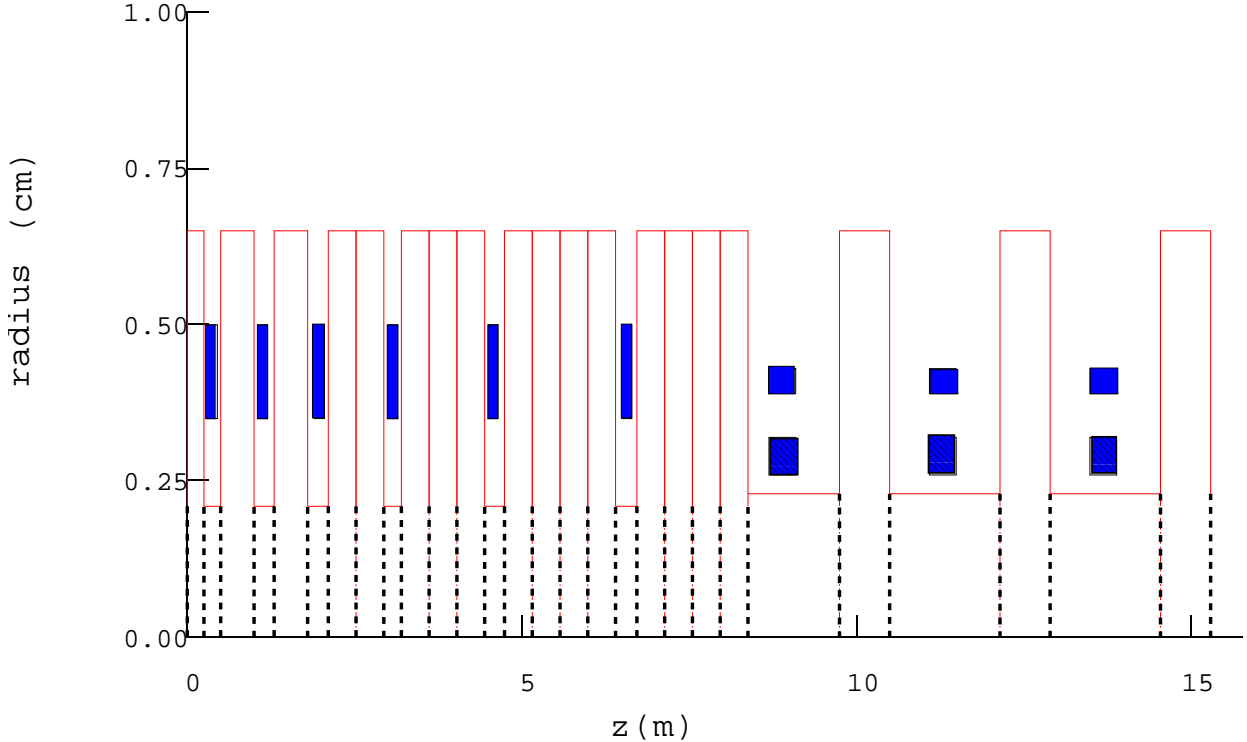}
  \caption{(Color) Matching section from cooling to linac.  Blue rectangles are
    solenoids, red lines are rf cavity walls and dashed black lines are the
    windows of the cavities.  The last three cavities
    are superconducting, the remaining ones are room temperature.}
  \label{fig:acc:match}
\end{figure}
The cooling section has a beta function of around 0.8~m, whereas the
beginning of the acceleration linac has a beta function of around 2.7~m.  A
matching section is required to gradually change the beta functions from
one section to the other so as to avoid emittance growth and/or particle
loss.  Furthermore, the reduced acceptance of the longer cells in the
acceleration linac, as compared to the more compact cells of the cooling
section, necessitates that the acceleration linac start at an energy above
that of the cooling section (see Table~\ref{tab:acc:pars}); the matching 
section will thus also begin to increase the beam energy after the 
cooling.
  That matching section will consist of six cells similar to
those in the cooling channel, but with increasing lengths and numbers of 
cavities per cell, and three superconducting cells similar to the 
accelerating linac, but made shorter by the use of shorter, higher field 
focusing solenoids.   
Figure~\ref{fig:acc:match} shows a layout of the matching section. 
The current design for the matching section has about 15\% loss;
initial studies indicate that this may be due to performing the
matching at low instead of high amplitudes.  Initial attempts at
performing the longitudinal match at high amplitudes have eliminated
the losses longitudinally, but we have not yet done the matching for the 
transverse plane as well. 

\subsection{Low Energy Acceleration}

Based on preliminary cost considerations, we have
chosen not to use FFAGs below 5~GeV total energy. Therefore, we must
provide alternative acceleration up to that point. Similarly to what was
adopted in FS2, we use a
linac from the lowest energies to 1.5~GeV, followed by a recirculating
linear accelerator (RLA).

The linac parameters are strongly constrained by the transverse
acceptance.  In FS2 there were three types of cryomodules, containing
one, two, and four two-cell cavities, respectively. Because of our larger
acceptance requirements, the cryomodule-dimensions from FS2 would require the 
beam to have a
momentum of at least 420~MeV/c, 672~MeV/c, and 1783~MeV/c,
respectively. Note that the momentum for the first stage of the linac  is, 
already, much higher than the
average momentum in the cooling channel, which is about 220~MeV/c.
Thus, we need to make adjustments to the FS2 design to be able to
accelerate this larger beam.

\begin{figure}[tbhp!]
\includegraphics{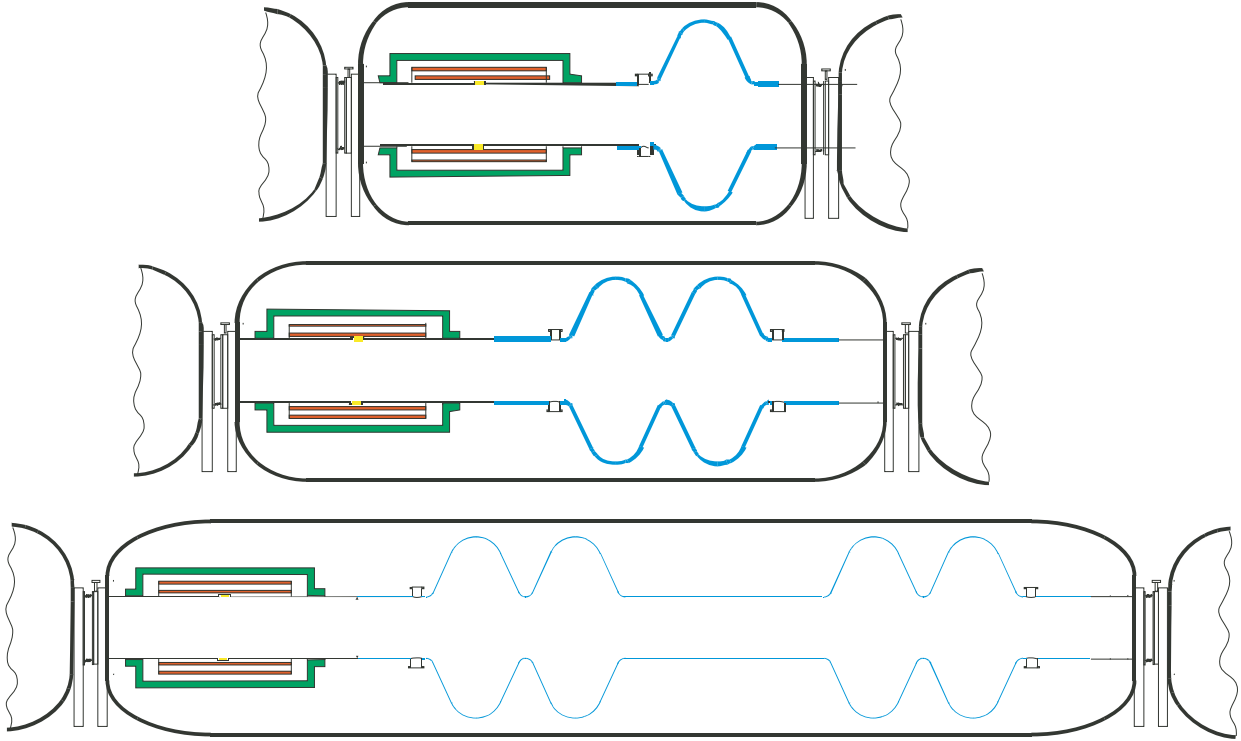}
\caption{(Color) Layouts of superconducting linac
pre-accelerator cryomodules. Blue lines are the SC walls of the cavities, 
solenoid coils are indicated in red, and the iron shielding is in green. The 
dimensions of the cryomodules are
shown in Table~\ref{tab:acc:cryo}, and Table~\ref{tab:acc:linac} summarizes 
parameters for the linac.}%
\label{fig:acc:cryomod}%
\end{figure}

\begin{table}[tbp]
  \caption{Linac cryomodule structure.
    Numbers are lengths in m.\label{tab:acc:cryo}}
  \begin{ruledtabular}
    \begin{tabular}{rlrlrl}
      \multicolumn{2}{c}{Cryostat I}&\multicolumn{2}{c}{Cryostat II}
      &\multicolumn{2}{c}{Cryostat III}\\
      \hline
      Drift&0.45&Drift&0.70&Drift&0.70\\
      Solenoid&1.00&Solenoid&1.00&Solenoid&1.00\\
      Drift&0.50&Drift&1.00&Drift&1.00\\
      Cavity&0.75&Cavity&1.50&Cavity&1.50\\
      Drift&0.30&Drift&0.80&Drift&1.50\\
      \cline{1-2}\cline{3-4}
      Total&3.00&Total&5.00&Cavity&1.50\\
      &&&&Drift&0.80\\
      \cline{5-6}
      &&&&Total&8.00
    \end{tabular}
  \end{ruledtabular}
\end{table}
In particular, to increase the acceptance, we must reduce the lengths
of the cryomodules.  We first employ a very short cryomodule 
using a single one-cell cavity as opposed to the two-cell cavities used in
all of the FS2 cryomodules.  Not only does this shorten the cavity itself, it
also eliminates one of the input couplers.  We also eliminate some of
the drift space in the cryomodule.  This is possible since we now consider
it acceptable 
to run the cavities with up to 0.1~T on them~\cite{Ono99}, provided the cavities 
are cooled down before the magnets
are powered.  The field profile of the solenoids shown in FS2
indicates that the iron shield on the solenoids is sufficient to bring
the field down to that level, even immediately adjacent to the solenoid
shield.  Together, these changes reduce the total length for the first
module type to only 3~m.  
Table~\ref{tab:acc:cryo} shows the dimensions of the cryostats we will
use and Fig.~\ref{fig:acc:cryomod} depicts all three of them.

\begin{table}[tbp]
\caption{Linac cryomodule parameters.\label{tab:acc:linac}}
\begin{ruledtabular}
\begin{tabular}{lrrr}
&{Cryo I}&{Cryo II}&{Cryo III}\\
\hline
      Length (m)&3.00&5.00&8.00\\
      Number of modules&12&18&22\\
      Cells per cavity&1&2&2\\
      Cavities per module&1&1&2\\
      Maximum energy gain per cavity (MeV)&11.2&22.4&22.4\\
      Cavity rf frequency (MHz)&201.25&201.25&201.25\\
      Solenoid length (m)&1&1&1\\
      Max Solenoid field (T)&1.5&1.8&4.0
\end{tabular}
\end{ruledtabular}
\end{table}

\begin{figure}
  \centering
  \includegraphics[width=0.75\linewidth]{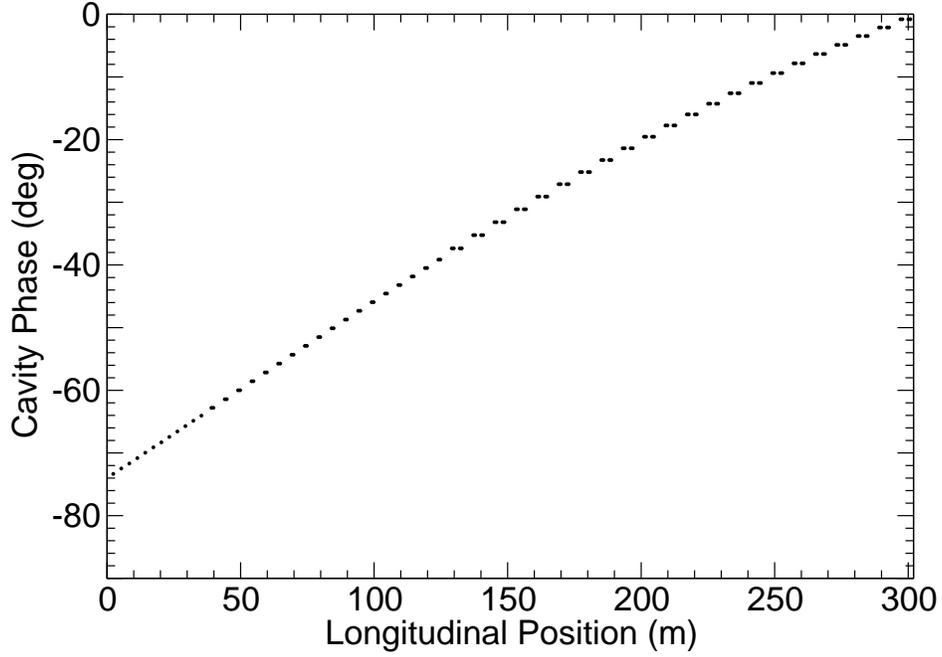}
  \caption{Cavity phase vs.\ position along the linac.}
  \label{fig:acc:phase}
\end{figure}
Table~\ref{tab:acc:linac} summarizes parameters for the
linac.  The phases of the cavities in the linac will be varied
approximately linearly with length from about $-73^\circ$ at the
beginning of the linac to $0^\circ$ at the end, as shown in
Fig.~\ref{fig:acc:phase}.
\subsection{RLAs}

\begin{figure}[tbp]
\includegraphics[width=\columnwidth]{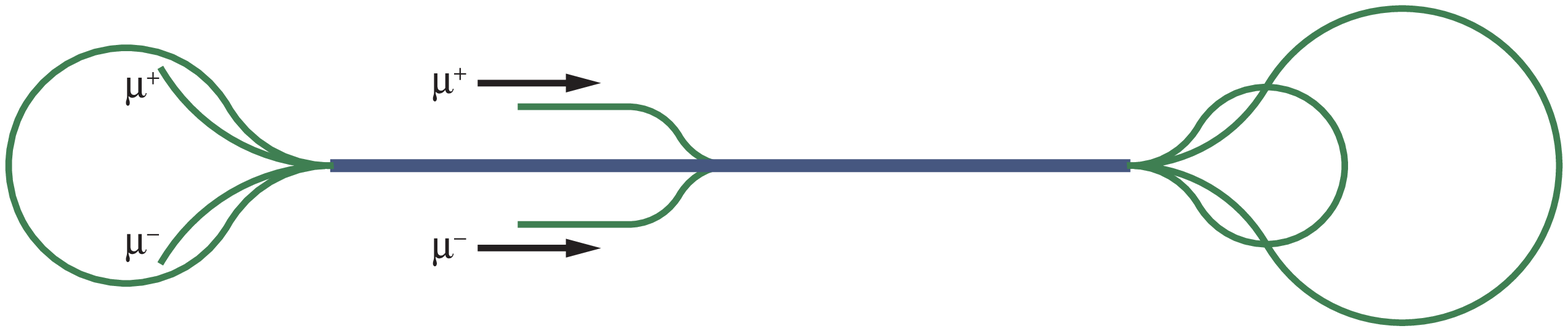}
\caption{(Color) Dogbone (top) and racetrack (bottom) layout for the RLA.} 
\label{fig:acc:layout}
\end{figure}

\begin{figure}[tbp]
\includegraphics[width=\columnwidth]{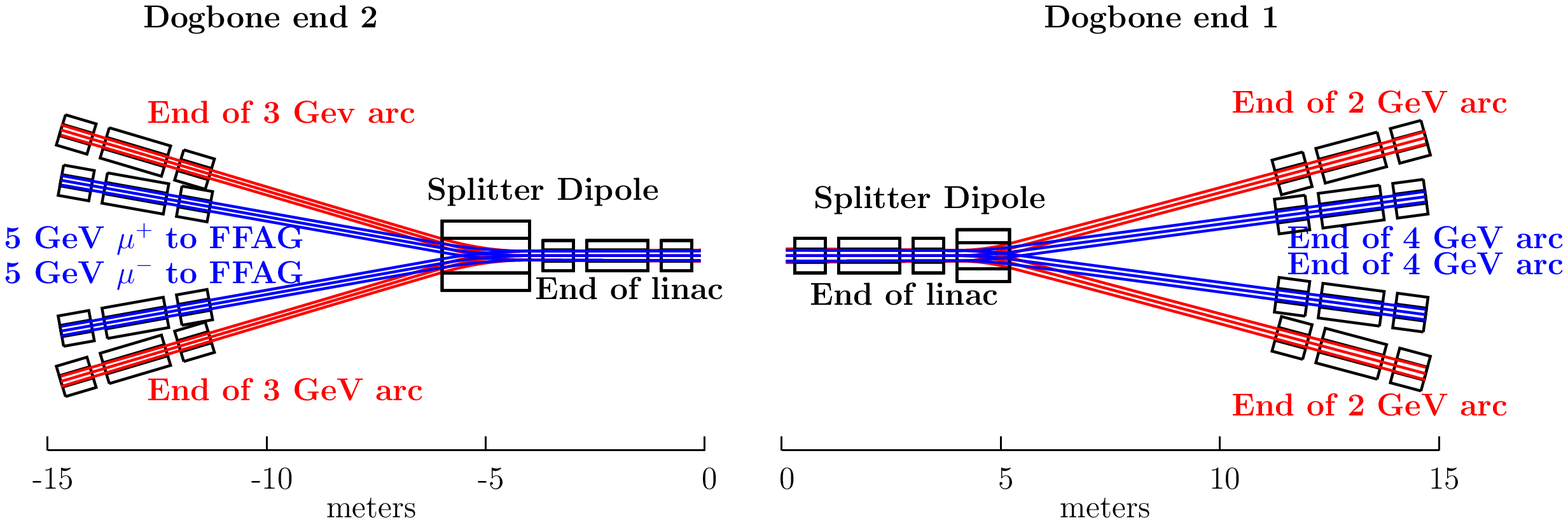}
\caption{(Color) Dogbone  RLA Switch Yards.} 
\label{fig:acc:switchyards}
\end{figure}

Compared with FS2, we are injecting into the RLA at a lower energy
and are accelerating over a much smaller energy range.  These features make
it more difficult to have a large number of turns in the RLA.  To mitigate this, 
we choose a dogbone layout for the RLA~\cite{pac01:3323}.  For a given amount of 
installed rf, the dogbone layout
has twice the energy separation of the racetrack layout at the
spreaders and recombiners (see Fig.~\ref{fig:acc:switchyards}), making the 
switchyard much easier and
allowing more passes through the linac. 

\begin{figure}[tbp]
  \centering
  \includegraphics[width=\columnwidth]{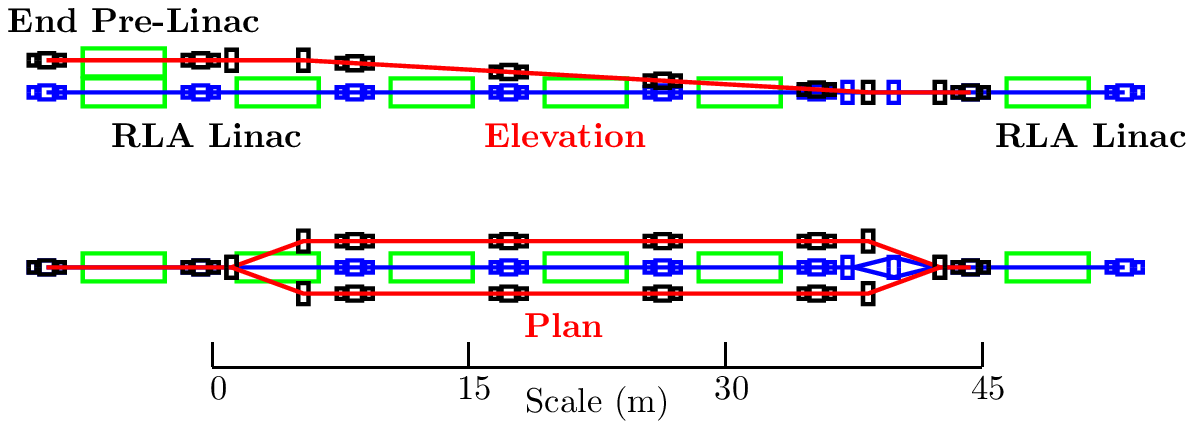}
  \caption{(Color) Injection chicanes. Magnets in black, rf in green, RLA
    linac in blue, and injection lines in red.}
  \label{fig:acc:chicanes}
\end{figure}

One disadvantage of the dogbone layout is that, because of the longer
linac and the very low injection energy, there is a significant
phase shift of the reference particle with respect to the cavity
phases along the length of the linac in the first pass, relative to later 
passes.  
 To reduce this effect, we inject into the center of the linac as shown
in Fig.~\ref{fig:acc:layout}. 
 This injection is accomplished with a chicane similar to that used for
injection in FS2, but here, to inject both signs, there are two chicanes,
one on either side of the linac (see Fig.~\ref{fig:acc:chicanes}). The start
of the chicanes  is the point at which the particles with differing charges
are first separated. To avoid this point overlapping the earlier part of
the linac, the chicanes are tilted slightly upwards.

\begin{figure}[tbp]
  \centering
  \includegraphics[width=\columnwidth]{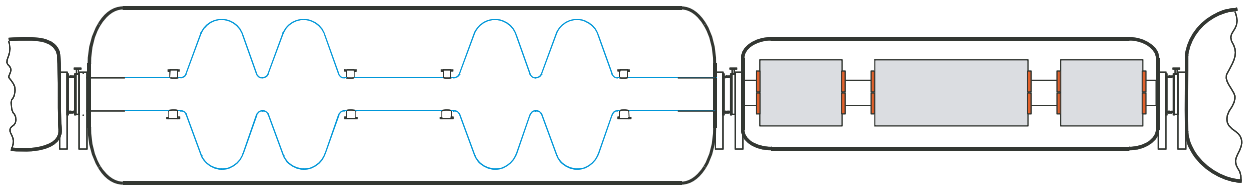}
  \caption{(Color) Dogbone linac cell.} 
  \label{fig:acc:doglinac}
\end{figure}
In the dogbone RLA we have just over 1~GeV of linac, and we make
three and a half passes through that linac to accelerate from a total
energy of 1.5~GeV to 5~GeV.  The RLA linac consists of 11~m long cells
with two 2-cell cavities per cell, and quadrupole triplet focusing, as shown in
Fig.~\ref{fig:acc:doglinac}.  The cavities are the 30~cm aperture
cavities assumed in FS2, as opposed to the 46~cm aperture cavities that were
used in the linac that accelerated up to 1.5~GeV; this should permit a
somewhat higher gradient (17~MV/m rather than  15~MV/m). 

The arcs will also use quadrupole triplet focusing, with a 90$^\circ$ phase 
advance
per cell in both planes, in order to cancel some chromatic effects. 
Both the quadrupoles and the dipoles in the arc and linac lattices
will have 1~T maximum field at the coils, and can be warm magnets. 

\begin{figure}[bthp!]
\includegraphics[width=\textwidth]{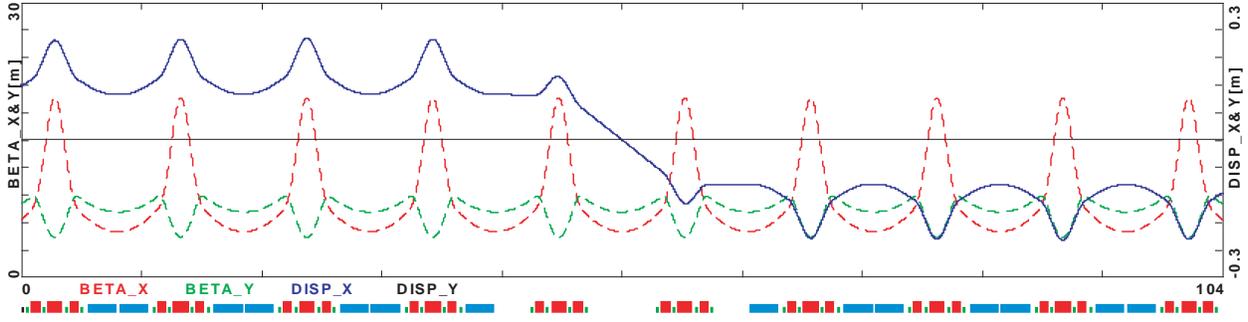}
\caption{(Color) A section of the dogbone arc where the bend changes direction,
    showing the dispersion (solid) and beta functions (dashed).} 
\label{fig:acc:disp}
\end{figure}
Since the dogbone arc changes its direction of bend twice in each arc,
dispersion matching must be handled carefully.  This is done
straightforwardly  by having
a 90$^\circ$ phase advance per cell, and removing the dipoles from two
consecutive cells.  This will cause the dispersion to switch to the
other sign as desired, as shown in Fig.~\ref{fig:acc:disp}. Matching of 
off-momentum particles is controlled using sextupoles.

\subsection{FFAGs\label{sec5-sub2-a}}
\begin{table}[tbp]
\caption{Parameters for FFAG lattices. See Fig.~\ref{fig:acc:ffaggeom} to
understand the signs of the parameters.}
\label{tab:acc:ffag}%
\begin{ruledtabular}
    \begin{tabular}{lrrrr}
      Maximum energy gain per cavity (MeV)&\multicolumn{4}{c}{7.5}\\
      Stored energy per cavity (J)&\multicolumn{4}{c}{368}\\
      Cells without cavities&\multicolumn{4}{c}{8}\\
      RF drift length (m)&\multicolumn{4}{c}{2}\\
      Drift length between quadrupoles (m)&\multicolumn{4}{c}{0.5}\\
      Initial total energy (GeV)&\multicolumn{2}{c}{5}&\multicolumn{2}{c}{10}\\
      Final total energy (GeV)&\multicolumn{2}{c}{10}&\multicolumn{2}{c}{20}\\
      Number of cells&\multicolumn{2}{c}{90}&\multicolumn{2}{c}{105}\\
      Magnet type&\multicolumn{1}{c}{Defocusing}&\multicolumn{1}{c}{Focusing}&
      \multicolumn{1}{c}{Defocusing}&\multicolumn{1}{c}{Focusing}\\
      Magnet length (m)&1.612338&1.065600&1.762347&1.275747\\
      Reference orbit radius of curvature (m)&15.2740&-59.6174&18.4002&-
70.9958\\
      Magnet center offset from reference orbit (mm)&-1.573&7.667&1.148&8.745\\
      Magnet aperture radius (cm)&14.0916&15.2628&10.3756&12.6256\\
      Field on reference orbit (T)&1.63774&-0.41959&2.71917&-0.70474\\
      Field gradient (T/m)&-9.1883&8.1768&-15.4948&12.5874\\
    \end{tabular}
  \end{ruledtabular}
\end{table}
\begin{figure}[htbp]
\includegraphics[width=0.65\textwidth]{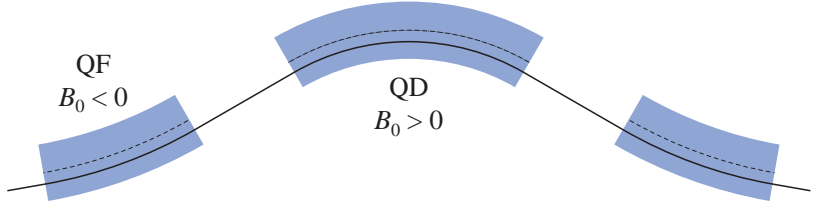}
\caption{(Color) Geometry of the Triplet Lattice. The ``magnet center offset 
from
reference orbit" listed in Table~\ref{tab:acc:ffag} is positive for both
magnets in this diagram.}
\label{fig:acc:ffaggeom}
\end{figure}

Once we reach 5~GeV, it appears to be more cost-effective to use FFAGs
rather than RLAs. This conclusion is based on applying a procedure
for producing minimum-cost FFAG designs~\cite{muc-309,jsb-cost} and comparing
the resulting costs to those from FS2. FFAG designs for accelerating from 5
 to 10~GeV and from 10 to 20~GeV 
  are given in Table~\ref{tab:acc:ffag}.  The lattices
consist exclusively of combined-function triplet cells with a drift
length sufficient to hold a single-cell 201.25~MHz superconducting rf
cavity (similar to the double-cell cavities from FS2).  The 2~m length
of the drift is needed to keep the fields on the cavity under
0.1~T~\cite{Ono99}. 

With the 1 MW beam intensity given in Table~\ref{tab:acc:pars}, and both
signs of muons, about 16\% of the stored energy will be extracted from
the cavities in the 5--10~GeV FFAG, and about 27\% will be extracted
in the 10--20~GeV. While this may seem substantial, it is easily
handled. To keep the average voltage sufficient to accelerate over the
desired range, 7.5~MV,  one need only to increase the initial voltage to 
7.8~MV for the 5--10~GeV FFAG and to 8.1~MV for the 10--20~GeV FFAG.  The
most important effect is a
differential acceleration between the head and tail of the bunch
train, which is only about 1\% for both cases. This should be at least
partially correctable by a phase offset between the cavity and the
bunch train and, in any case, is substantially smaller than the energy
spread in a single bunch. 

\begin{table}[tbp]
  \caption{Parameters for FFAG injection and extracting kickers.} 
  \label{tab:acc:injext}
  \centering
  \begin{ruledtabular}
    \begin{tabular}{lcccc}
      Energy (GeV)&5&10&10&20\\
      Type&Inject&Extract&Inject&Extract\\
      Length (m)&1.5&1.5&1.5&1.5\\
      Kick field (T)&0.37&0.51&0.78&0.58\\
      Maximum field at the coils (T)&3.6&2.6&4.2&5.6\\
      Vertical aperture (cm)&10&10&7.6&7.6\\
      Horizontal aperture (cm)&25&25&19.5&19.5\\
      Current (kA)&44&60&71&53\\
      Supply voltage (kV)&$\pm58$&$\pm60$&$\pm52$&$\pm48$\\
      Rise/fall time (ns)&640&950&875&1270\\
      Pulse length (ns)&300&300&300&300\\
      Stored energy (J)&850&1620&2280&1260\\
    \end{tabular}
  \end{ruledtabular}
\end{table}
\begin{figure}[tbp]
  \centering
  \includegraphics[width=\columnwidth]{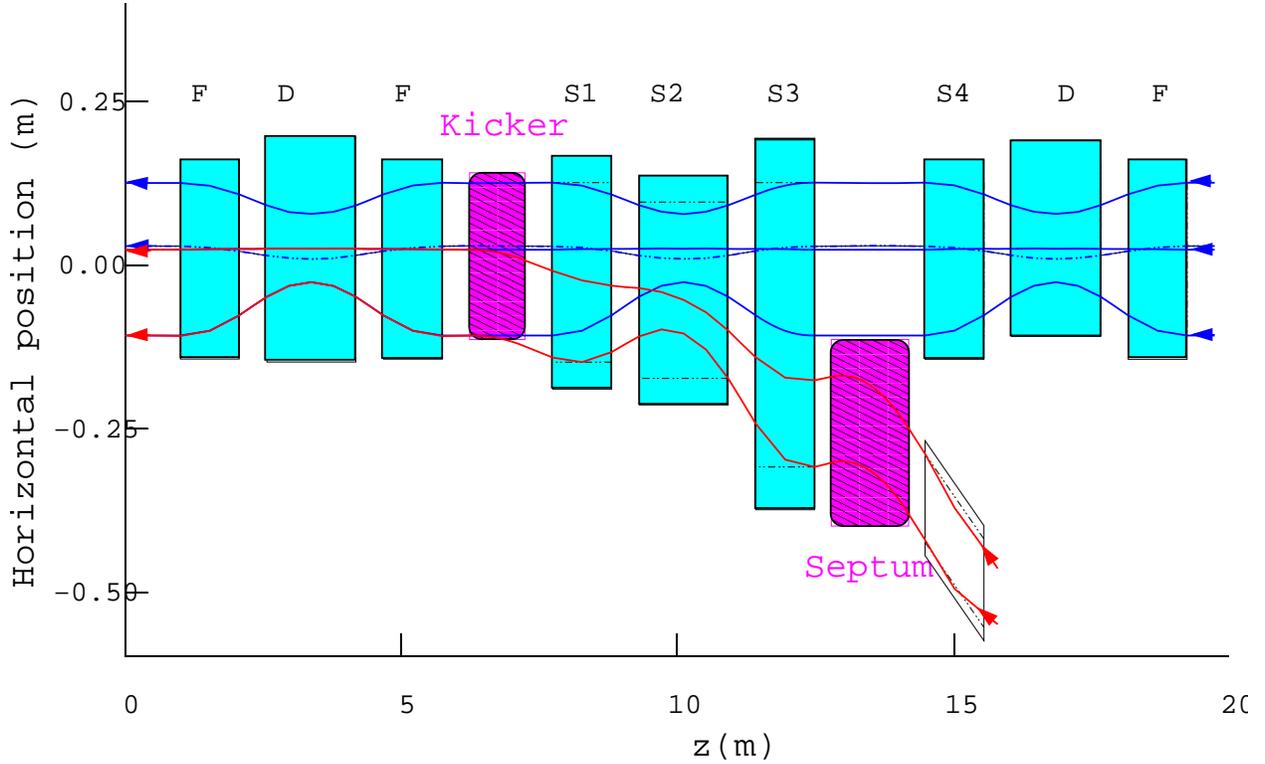}
  \caption{(Color) Injection into the FFAG. S1 to S4 are special injection 
region dipoles.} 
  \label{fig:acc:inj}
\end{figure}
\begin{figure}[tbp]
  \centering
  \includegraphics[width=\columnwidth]{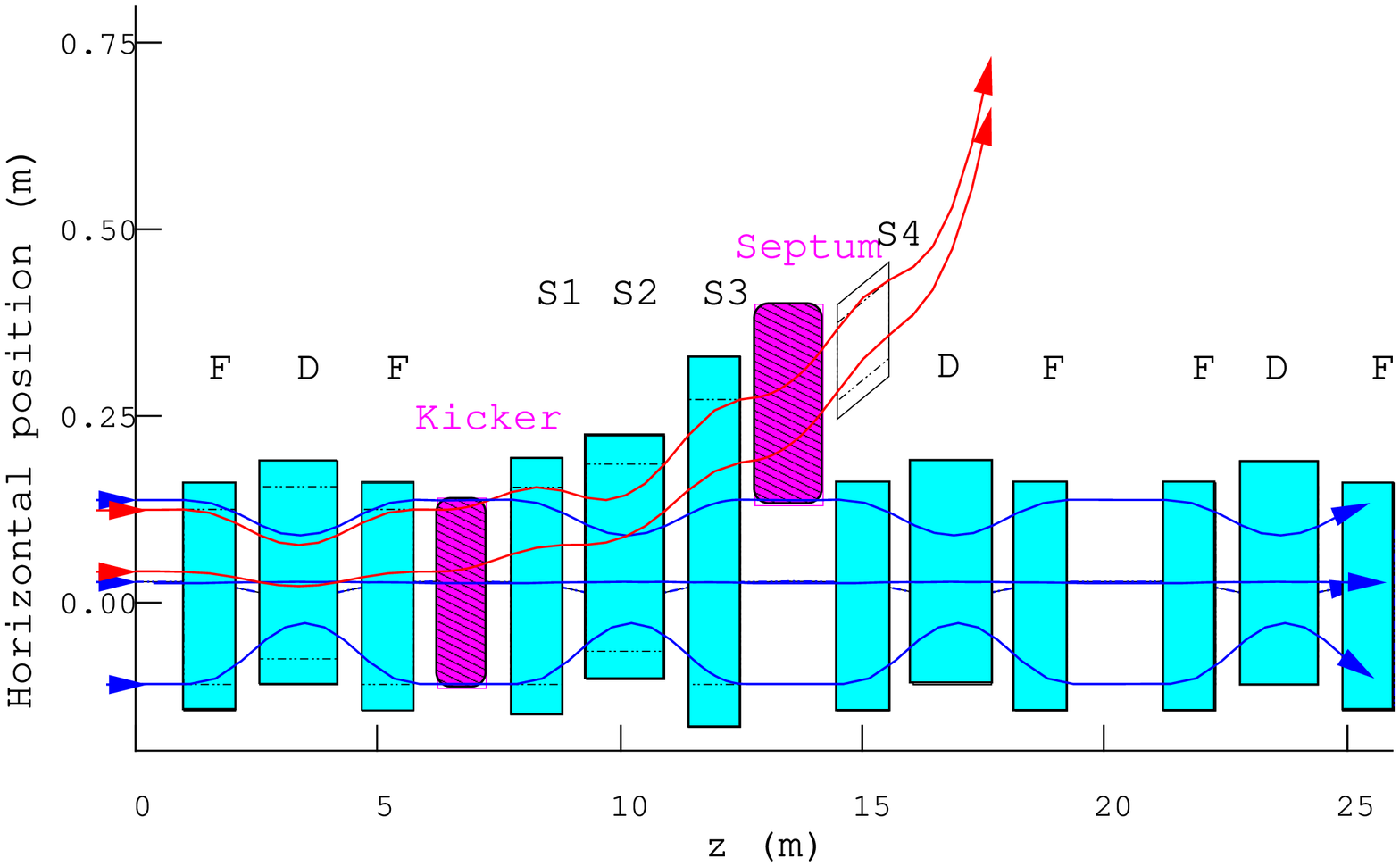}
  \caption{(Color) Extraction out of the FFAG. S1 to S4 are special extraction 
region dipoles.} 
  \label{fig:acc:ext}
\end{figure}
One of the biggest challenges for the FFAGs is injection and
extraction.  Table~\ref{tab:acc:injext} gives the parameters required
for injection and extraction kickers.  The stored energy in the kicker
is high, but is similar to that found in induction linac cells.  The
rise times and voltages are also similar to those in induction linacs.  These 
parameters assume
that injection occurs from the inside of the FFAG.  This is preferred
since the beam will be near the inside of the FFAG at the lowest
energies.  Figures \ref{fig:acc:inj} and \ref{fig:acc:ext} show the
injection and extraction layout.  The magnets near the kickers and
septum must be modified to accommodate the injection and extraction
systems, but their effects will be kept as close as possible to those
of the other cells in the FFAG lattice to minimize the driving of
resonances. 

\subsection{FFAG Tracking Results}
Initial experience with FFAG lattices having  a linear midplane field profile
has shown them to have a good dynamic aperture at fixed energies. We are
careful to avoid single-cell linear resonances to prevent beam loss. 
However, since the tune is not constant, the
single-cell tune will pass through many nonlinear resonances. Nonlinearities
in the magnetic field due to end effects are capable of driving those
nonlinear resonances, and we must be sure that there is no beam loss and
minimal emittance growth because of this. Furthermore, there is the
potential to weakly drive multi-cell linear resonances because the changing
energy makes subsequent cells appear slightly different from each other. 
These effects can be studied through tracking. 

ICOOL~\cite{icool} is used for tracking for several reasons. It will
allow for a fairly arbitrary end-field description, it forces that
description to be consistent with Maxwell's equations, and it will track
accurately even when the lattice acceptances, beam sizes, and energy spread are 
all large. 

\begin{figure}[tbp]
\includegraphics[width=0.5\textwidth]{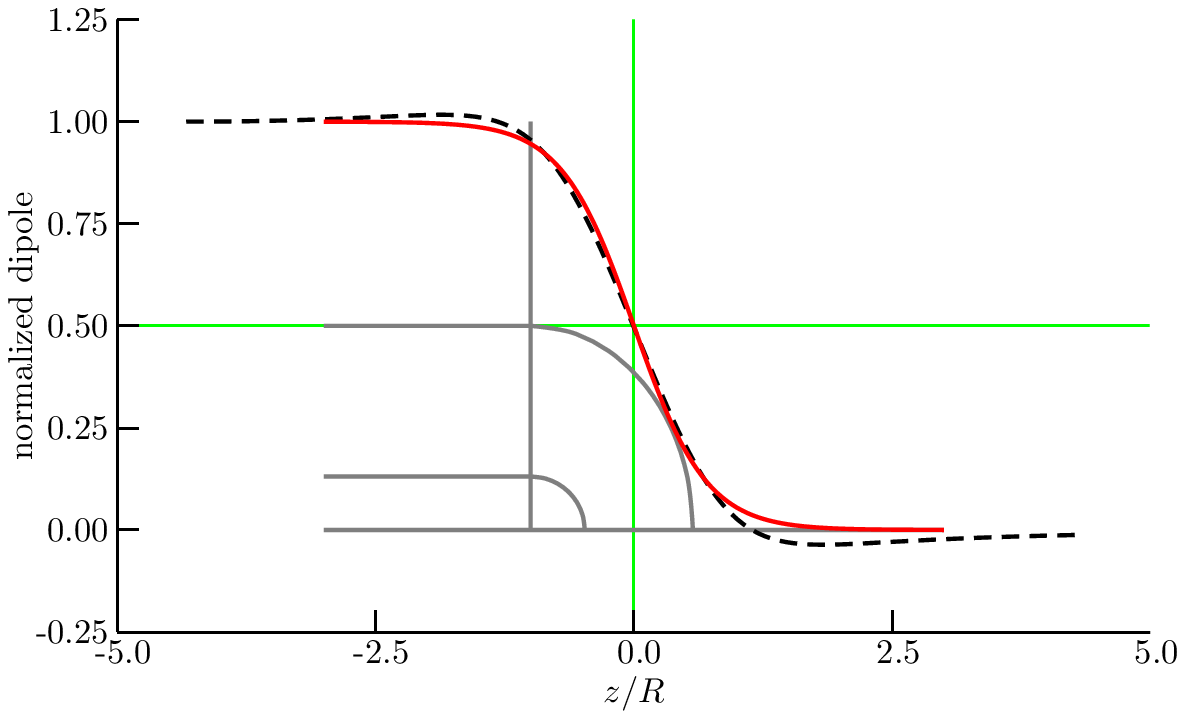}%
\includegraphics[width=0.5\textwidth]{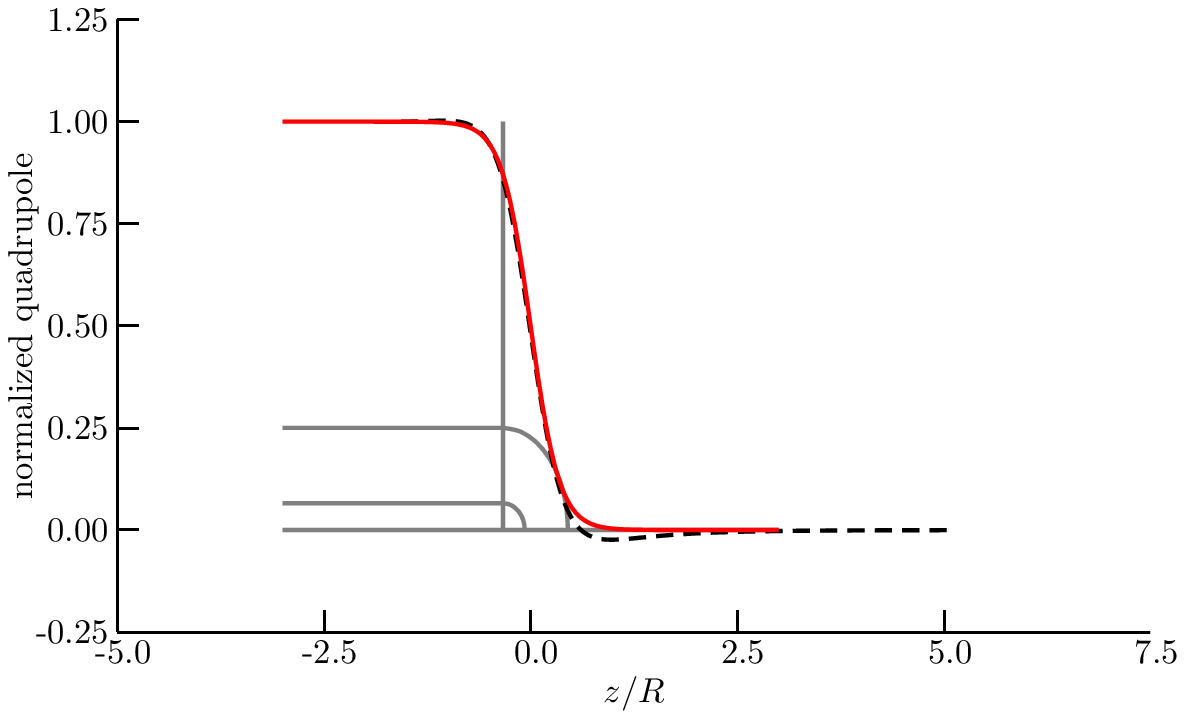}
\caption{(Color) Relative dipole field (left) and quadrupole field (right) near 
the
magnet end. The dashed line is the field from TOSCA, while the solid line is
our model.} 
\label{fig:acc:end0}
\end{figure}
\begin{figure}[tbp]
\centering
\includegraphics[width=0.65\textwidth]{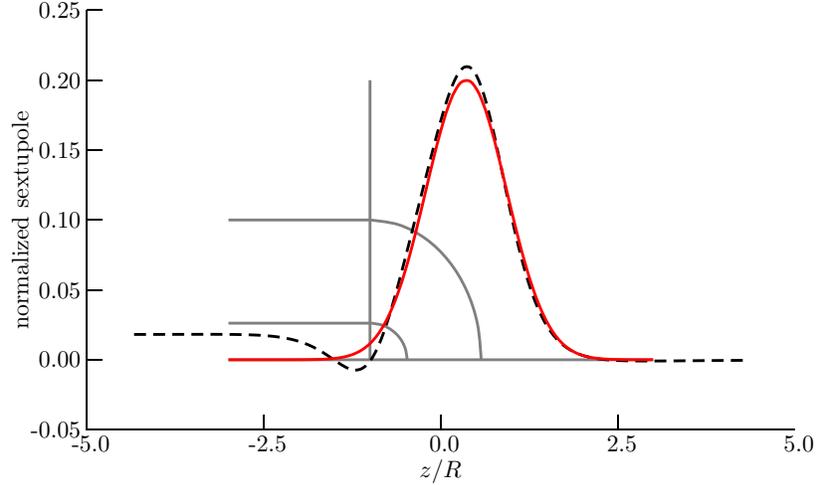}
\caption{(Color) Peak magnitude of the sextupole end field at radius $R$ (the 
magnet
aperture), divided by the dipole field. The dashed line is the field from
TOSCA, while the solid line is our model.} 
\label{fig:acc:end2}
\end{figure}

We begin by constructing a simple model of both a quadrupole and dipole $%
\cos\theta$-type magnet, without iron, using TOSCA~\cite{opera3d}. At the end
of the magnet, the field does not immediately drop to zero, but falls
gradually, as shown in Fig.~\ref{fig:acc:end0}. The end-field falloff in a
dipole or a quadrupole generates nonlinear fields, which ICOOL calculates. In
addition, there are higher-order multipoles generated by breaking the
magnet symmetry at the ends where the coils form closed loops. We use
TOSCA to compute the sextupole components that arise from this effect, as shown 
in Fig.~\ref{fig:acc:end2}, and include them in our computation. 

The TOSCA computation is done without iron, which leads to the overshoot in
the field values in Figs.~\ref{fig:acc:end0} and ~\ref{fig:acc:end2}. Iron in
the magnet will likely eliminate that overshoot. Thus, we approximate the
fields from TOSCA using functions without the overshoot. Fitting roughly to
the TOSCA results, the fields are approximated by 
\begin{equation}
\begin{gathered} B_0(z) = 
\dfrac{1}{2}B_{00}\left(1+\tanh{\dfrac{z}{0.7R}}\right),\qquad
B_1(z) = \dfrac{1}{2}B_{10}\left(1+\tanh{\dfrac{z}{0.35R}}\right)\\ B_2(z) =
-0.2B_{00}\exp\left[-\dfrac{1}{2} \left(\dfrac{z-0.36 R}{0.57
R}\right)^2\right], \end{gathered}
\end{equation}
where $R$ is the magnet aperture radius, $B_0(z)$ is the dipole field, $%
B_{00}$ is the dipole field in the center of the magnet, $B_1$ is the
quadrupole field, $B_{10}$ is the quadrupole field in the center of the magnet,
and $B_2$ is the maximum magnitude of the sextupole field at the radius $R$. 
These fitted functions are shown in red in  their corresponding plots in
Figs.~\ref{fig:acc:end0} and ~\ref{fig:acc:end2}. 


\begin{figure}[tbp]
\centering
\includegraphics[width=0.65\textwidth]{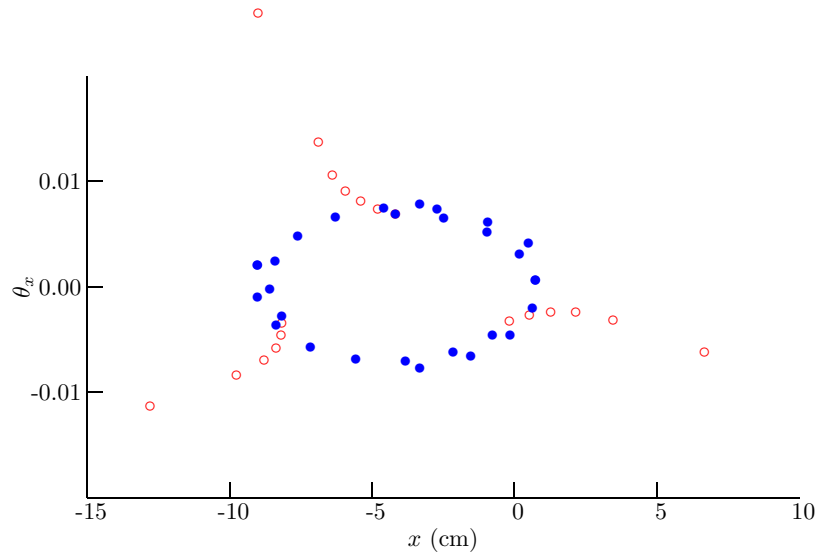}
\caption{(Color) Horizontal phase space from tracking at 5.1~GeV/c at the outer 
edge
of the acceptance. Open circles are without the body sextupole fields and
show a third-order resonance; filled circles are with the body sextupole
fields.} 
\label{fig:acc:res3}
\end{figure}
\begin{figure}[tbp]
\centering
\includegraphics[width=0.65\textwidth]{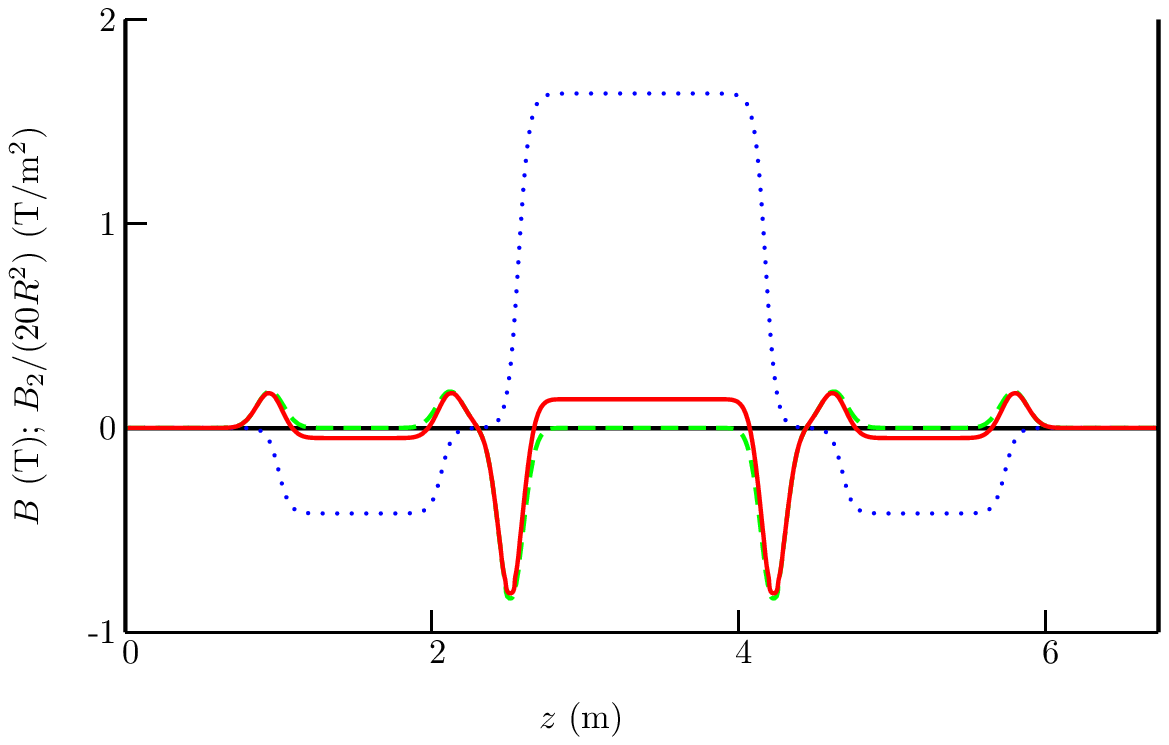}
\caption{(Color) Sextupole field components in the lattice used for
  tracking.  The dotted line is the dipole field, the dashed line is
  $B_2/(20R^2)$ with zero body sextupole field, and the solid line is
  with sufficient body sextupole field to eliminate the third-order
  resonance.} 
\label{fig:acc:sex}
\end{figure}
\begin{figure}[tbp]
\includegraphics[width=0.65\textwidth]{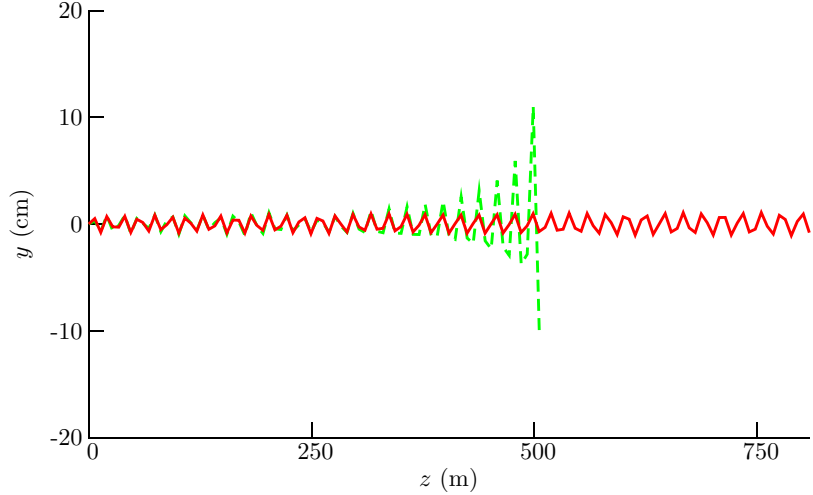}%
\caption{(Color) Tracking of a particle at the edge of the acceptance
  with uniform acceleration.  The dashed line is without
  any body sextupole, and the solid line is with the corrected body
  sextupole. 
} 
\label{fig:acc:trkacc}
\end{figure}

Injecting particles at the outer edge of the acceptance and tracking
through several cells indicated a large third-order resonance at
around 5.1~GeV/c, as shown in Fig.~\ref{fig:acc:res3}. This resonance
is presumably being driven by the sextupole fields at the magnet ends. 
With some experimentation, it
was found that if the integrated body sextupole was set to 68\% of the
integrated end sextupoles, (see Fig.~\ref{fig:acc:sex}), the resonance
was eliminated (also shown in Fig.~\ref{fig:acc:res3}). When
acceleration is included, one sees particle loss when accelerating
through the resonance if there is no body sextupole correcting the end
sextupoles, while there appears to be almost no loss with the body
correction included (see Fig.~\ref{fig:acc:trkacc}). 
If the body
correction is only partially included, there is significant emittance
growth. With these sextupole
corrections, we can uniformly accelerate over the entire 5--10~GeV
energy range without losing a high-amplitude particle or having its
amplitude grow by a large amount. 

\begin{figure}[tbp]
\includegraphics[width=0.5\textwidth]{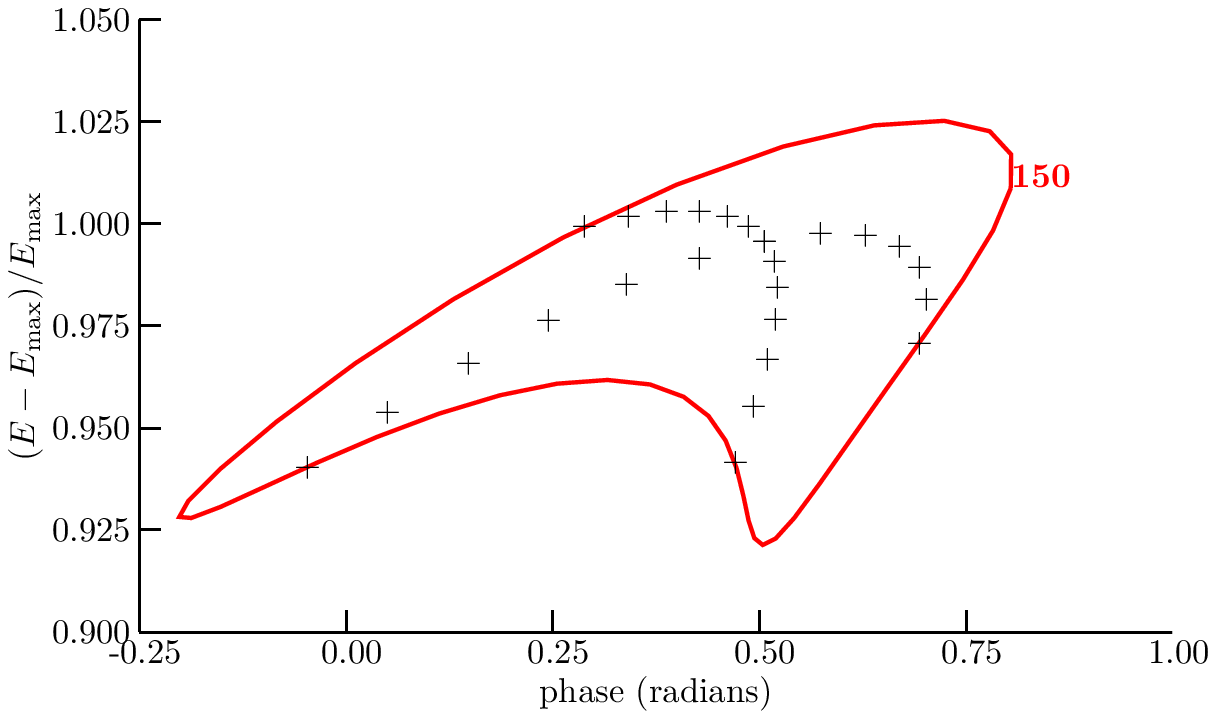}%
\includegraphics[width=0.5\textwidth]{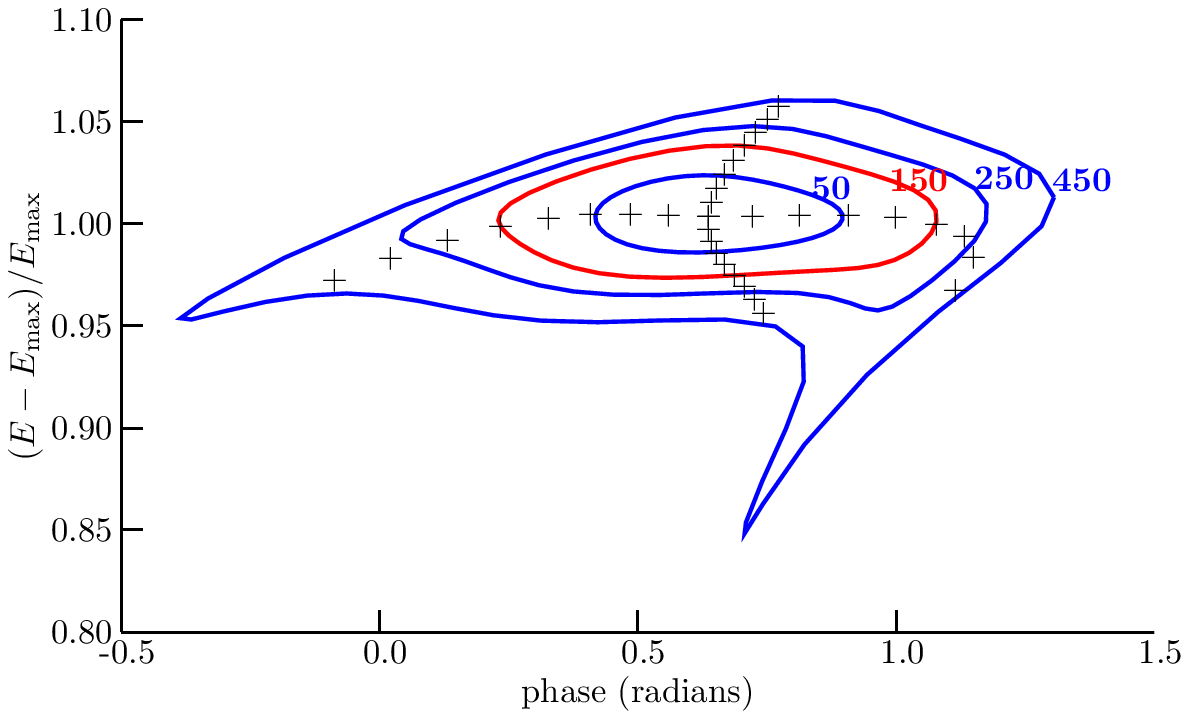}
\caption{(Color) Longitudinal tracking starting from an upright
  ellipse for the 5--10~GeV FFAG. On the left with only 201.25~MHz
  rf. On the right with third-harmonic rf having voltage equal to 2/9
  of the fundamental rf voltage.  Curves are labeled with their
  corresponding acceptance. Crosses for both cases started out as
  horizontal and vertical lines in phase space.} 
\label{fig:acc:longtrk}
\end{figure}

When tracking with rf is considered, the longitudinal dynamics
is complex~\cite{KoscielniakPAC03}. If one begins with an
upright ellipse, there is considerable emittance growth if only the
201.25~MHz rf is used (see Fig.~\ref{fig:acc:longtrk}). Adding a
third-harmonic rf considerably reduces the emittance growth, as shown
in Fig.~\ref{fig:acc:longtrk}. The amount of third-harmonic rf
required is substantial and that, combined with space considerations,
makes this alternative unattractive. An alternative that includes tilting
the initial ellipse in phase space, which also reduces the emittance growth, is 
being studied.

\subsection{Design of Combined-Function Superconducting Magnet for 
FFAGs\label{COM-FUNC-MAG}}
An initial design of a superconducting combined-function
(dipole--quadrupole) magnet has been developed~\cite{CaspiRef}.  
The work has been done for the defocusing magnet from the above design. 
  The parameters of this QD
combined-function magnet are shown in Table~\ref{qdcell}.
\begin{table}[htbp!] 
\caption{Parameters of the QD magnet: $L_0$ is the length of the long drift
  between the QF magnets; $L_q$ is the length of the short drift between QF
  and QD magnets; $X_0$ is the displacement of the center of the magnet
  from the reference orbit (see Fig.~\ref{fig:acc:ffaggeom}); $B_0$ is the vertical magnetic field at the 
reference orbit, and $B_1$ is the derivative of the vertical magnetic field at 
the reference orbit.}
\label{qdcell} 
\begin{ruledtabular}
\begin{tabular*}{10cm}{lc}
Initial energy, $E_{\text{min}}$ (GeV)& 10\\
Final energy, $E_{\text{max}}$ (GeV) &20\\
Long drift, $L_0$ (m) &2\\
Short drift, $L_q$ (m) &0.5\\
Type of magnet & QD\\
Length of reference orbit, $L$ (m)&1.762\\
Radius of curvature, $r$ (m)& 18.4\\
Displacement, $X_0$ (mm)&1.148\\
Radius of the magnet bore, $R$ (cm)&10.3756\\
Vertical magnetic field, $B_0$ (T)&2.7192\\
Gradient, $B_1$ (T/m)&-15.495\\
\end{tabular*}
\end{ruledtabular}
\end{table}

\begin{figure}[hbtp!] 
\includegraphics*[width=4in]{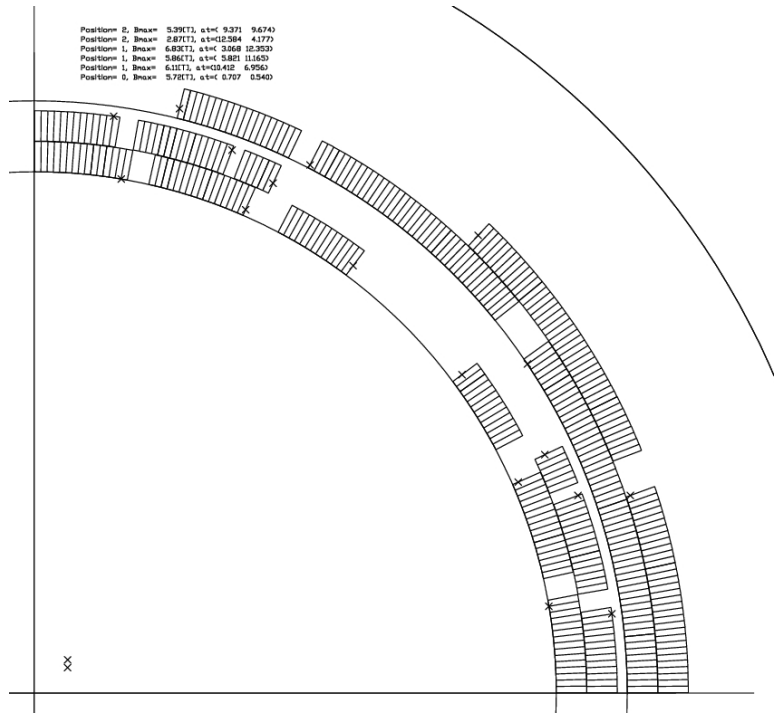}
\caption{First quadrant of the combined-function magnet cross section.} 
\label{first-quad}
\end{figure}

The magnet design is based on a cosine-theta configuration with two double
layers for each function. A cross section for one quadrant is shown in  
Fig.~\ref{first-quad}. The quadrupole coil is located within the dipole
coil  and both coils are assembled using key-and-bladder technology. All
coils are made with the same Nb--Ti cable capable of generating the
operating dipole field and gradient with about the same current of 1800~A. 
 A single power supply is thus possible with a bit of fine tuning. 
 The maximum central dipole field and gradient at short sample are 4.1~T and 
26~T/m, as compared 
with the requirements of 2.7~T and 15.4~T/m, respectively. At this early
design stage, excess margin is left for safety and perhaps a field rise in the 
magnet end region. The maximum azimuthal forces required for magnet 
pre-stress are of the order of 1~MN/m (assuming maximum safety). 
 The conductor strand size and cable parameters common to both dipole and 
quadrupole are listed in Table~\ref{nbti}. 
\begin{table}[htbp!] 
\caption{Nb--Ti conductor for dipole and quadrupole coils.\label{nbti}}
\begin{ruledtabular}
\begin{tabular}{lc}
Strand diameter (mm) &0.6477\\
Cable width, bare (mm) & 6.4\\
Cable thickness, insulated (mm) &1.35\\
Keystone angle  (deg.) &0.6814\\
Conductor type &Nb--Ti\\
Cu:SC ratio  & 1.8:1\\
Current density (at 5~T, 4.2~K) (A/{mm}$^2$)& 2850\\
Number of strands& 20\\
\end{tabular}
\end{ruledtabular}
\end{table} 

The initial cross sections of both dipole and quadrupole were designed to
give less than 0.01 units of  systematic 
multipole errors at a radius of 70~mm. It is straightforward to readjust
the design to cancel the end-field multipoles. 

\begin{figure}[hbpt!] 
\includegraphics[width=4in]{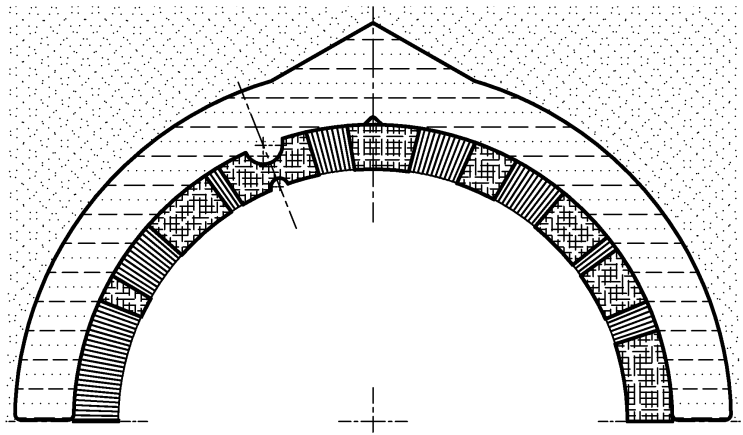}
\caption{ Expanded view of the cross section of the superconducting
  combined function magnet used in the 50~GeV proton beamline for the
  J-PARC neutrino experiment~\cite{JPARC-magnet1}.} 
\label{newmag}
\end{figure}

An alternative concept would be
to use a single dipole-like design with laterally displaced poles (see 
Fig.~\ref{newmag}) as
discussed in Refs.~\cite{JPARC-magnet1,JPARC-magnet2,JPARC-magnet3}.

\section{Muon Storage Ring and Performance\label{ring}}

The storage ring in this study is assumed to be essentially identical to
that in FS2. However, injection will be required in two opposite directions
for the two differing signs. The injection lines must be designed such that
when the train of one sign is traveling towards the detector, the train of
the other sign is moving away from the detector. In this way the neutrinos
of opposite kind arrive at well separated times, and the experiment can
analyze their reactions separately. Another difference is that in this
study, both straight sections must be designed with very high betas, so
that the neutrino beams of both types are well collimated. Finally, it must
be noted that the total energy deposited in the ring is doubled by the
presence of equally intense muon beams, but now of  two signs.

Losses are summarized in table~\ref{tab:ring:losses}. We define $\eta$ to be 
the probability that a muon makes it successfully into the storage ring.
The number of decays $N_\mu$, of each sign, injected into the storage ring in a 
$10^{7}$ second year is given by:
$$
N_\mu=10^{7}~f~N_p~\mu /p~\eta\approx 
10^{7}~\times~15~\times~(17\times
10^{12})~\times~0.17~\times~0.67\approx 2.9\times 10^{20}.
$$
  We define $\eta_{\rm straight}$ to be the length of the straight section 
pointing to the detector divided by the circumference of the storage ring. The 
number of decays $N_\nu$, of both signs, in the storage ring, decaying towards 
the detector, in a $10^{7}$ second year is given by:
$$
N_\nu~= 2~N_\mu~\eta_{\rm straight}\approx 2\times 2.9\times 10^{20}\times
0.35 \approx 2.0\times 10^{20}.
$$

This is a factor of two greater than that reported in FS2.
Note that if the proton driver power could be raised to 5~MW (4~MW has been
discussed in a further upgrade of the BNL AGS), then the number of
neutrinos per year would match the high performance goal ($10^{21}$)
suggested at the first  NuFact Workshop in Lyon, France~\cite{nufact99}.
\begin{table}[tbp]
\caption{Particle losses from cooling to storage ring}
\label{tab:ring:losses}
\begin{ruledtabular}
\begin{tabular}{lc}
&loss \%\\
\hline
Match into linac &15\\
Linac&6.0 \\
RLA  &6.1 \\
FFAG 1&5.1 \\
FFAG 2&6.5 \\
\hline
Total losses & 33 \\
\end{tabular}
\end{ruledtabular}
\end{table}

\section{Required R\&D\label{RandD}}
As should be clear from the design descriptions, the muon-based Neutrino Factory is a
demanding project. The machine makes use of novel components and techniques that are, in
some cases, at or beyond the state of the art. For this reason, it is critical
that R\&D efforts to study these matters be carried out. 


Each of the major systems has significant issues that must be addressed by
R\&D activities~\cite{aps-study}. Component specifications need to be verified. For example,
the cooling channel assumes a normal conducting rf (NCRF) cavity gradient of
15~MV/m at 201.25~MHz in substantial magnetic fields. Observations of
breakdown in 805 MHz cavities have shown~\cite{DarkCurrentnote} serious reductions of
attained rf gradients when the cavity is operated in a field. It is not
clear how to scale these observations to the 201.25~MHz case, so
experimental tests are urgently needed. If the required gradients cannot be
achieved in the specified magnetic  fields, then significant redesign will
be  needed.

The acceleration section demands high gradients
from superconducting rf (SCRF) cavities at this frequency; our requirements are somewhat beyond the performance reached to date for
cavities in this frequency range~\cite{newscottref}.

Development and testing of efficient high-power rf sources at a frequency near 200~MHz is also needed. 

The ability of the target to withstand a proton beam power at 1~MW and above must be confirmed. 

 Finally, an ionization
cooling experiment should be undertaken to validate the implementation and
performance of the cooling channel, and to confirm that our simulations of the
cooling process are accurate.

\section{Cost Estimate: Assumptions and Algorithm\label{sec9}}
For this study (ST2B) substantial effort has been directed at simplifying the 
design and thus hopefully reducing the 
costs of three major components of a neutrino factory: phase 
rotation, cooling, and the higher energy part of the muon
acceleration. A preliminary comparison with FS2, which contained detailed cost estimates with significant engineering 
input,  shows that a great deal of progress has been achieved.  
Starting from the FS2 work breakdown schedule,  we derived element costs per 
unit length, 
integral rf voltage, or net acceleration. For all but the final FFAG 
acceleration, these costs were then 
applied to the ST2B parameters after scaling for magnetic fields,
radii, stored energy, rf gradient, etc. For the FFAG costs a new cost algorithm 
had to be developed~\cite{jsb-cost}. Further details on the costing algorithms 
and their 
application to this new design can be found in a recent reference~\cite{MC-322}.

There are a number of reasons why we believe this new design should be 
significantly less expensive than the previous one described in FS2. For 
phase rotation we have replaced 260~m of expensive induction linacs with
54~m of more conventional rf cavities. The 108~m long cooling channel in
FS2 was 
replaced with an 80~m channel. The new cooling channel uses a simplified 
magnetic lattice with reduced peak solenoid fields; it also replaces liquid 
hydrogen absorbers with solid LiH absorbers. For acceleration to 20~GeV we 
have made use of recent advances in the development of non-scaling FFAG accelerators 
to replace most of the recirculating linear accelerator used in FS2. There are 
of course some changes in the new design, such as the adopted increase in 
transverse acceptance in the accelerators, which will increase the costs over 
FS2. However, our examination shows that the design changes should lead to an 
overall reduction in costs.

A summary of the preliminary estimates for the percentage cost reductions for 
the ST2B neutrino factory design is presented in
Table~\ref{s2atabsys}. It is most likely that a proton driver will first be 
built in conjunction with a neutrino super-beam experiment, so we begin the 
neutrino factory systems with the target and capture section. Excluding
the proton driver the new design should cost $\approx 35\%$ less than the FS2 
design.

\begin{table}[hpbt!]
\caption{Preliminary cost comparison between FS2 and ST2B.}
\label{s2atabsys}
\begin{ruledtabular}
\begin{tabular}{lc}
System & Reduced Cost \%\\ 
\hline
Target, capture, 18 m drift&98\\
Bunching and Phase Rotation & 39\\
Cooling                   & 60\\
Acceleration           &77\\
Ring                   & 100\\
\hline
Total without driver \& controls   &65\\ 
\end{tabular}

\end{ruledtabular}
\end{table}

\section{Summary\label{sec7}}
A new type of facility has been proposed that could have a tremendous 
impact on future neutrino experiments---the Neutrino Factory. 
In contrast to conventional muon~neutrino beams, 
the Neutrino Factory  would provide a 
source of electron~neutrinos $(\nu_e)$ and antineutrinos $(\bar{\nu}_e)$ 
with very low systematic uncertainties on the  
beam fluxes and spectra. The experimental signature for 
$\nu_e \to \nu_\mu$ transitions is extremely clean, with 
very low background rates. Hence, Neutrino Factories  
would enable very sensitive oscillation measurements to be made. 

A substantial Neutrino Factory R\&D effort has been ongoing in the U.S. and
elsewhere over the last few years, and significant progress has been made
towards optimizing the design, developing and testing the required accelerator
components, and significantly reducing the cost.

The novel facility described here represents a significant 
improvement over previous designs. New ideas in bunching, phase rotation, 
and ionization cooling have been incorporated into the design of the front
end, which now captures both muon signs simultaneously. The non-scaling FFAG acceleration concept has been further developed
and used for accelerating the muons up to the 20~GeV design energy. The
performance of the new system equals that of the earlier FS2, for each of
two neutrino states ($\nu$ and $\bar{\nu}$) that are generated essentially
simultaneously. The performance is thus effectively twice that of FS2. At
the same time, the facility is simpler than that in FS2 and of the order of
35\% less costly.

R\&D is also continuing to confirm needed component performance and establish the physical concepts used. Continued optimization is ongoing, and is expected to further improve performance and reduce the cost.


\begin{acknowledgments}
We would like to thank C. Johnstone, S.~Geer, and Y.~Fukui for helpful discussions.
 This research was supported by the U.S. Department of Energy under
Contract {Nos. DE-AC02-98CH10886}, {DE-AC02-76CH03000}, and {DE-AC03-76SF00098}.
\end{acknowledgments}
\bibliography{PR-STAB}

\begin{thebibliography}{40}
\expandafter\ifx\csname natexlab\endcsname\relax\def\natexlab#1{#1}\fi
\expandafter\ifx\csname bibnamefont\endcsname\relax
  \def\bibnamefont#1{#1}\fi
\expandafter\ifx\csname bibfnamefont\endcsname\relax
  \def\bibfnamefont#1{#1}\fi
\expandafter\ifx\csname citenamefont\endcsname\relax
  \def\citenamefont#1{#1}\fi
\expandafter\ifx\csname url\endcsname\relax
  \def\url#1{\texttt{#1}}\fi
\expandafter\ifx\csname urlprefix\endcsname\relax\def\urlprefix{URL }\fi
\providecommand{\bibinfo}[2]{#2}
\providecommand{\eprint}[2][]{\url{#2}}

\bibitem[{\citenamefont{{S.~Ozaki, R.~Palmer, M.~Zisman, and J.~Gallardo,
  eds.}}(2001)}]{fs2}
\bibinfo{author}{\bibnamefont{{S.~Ozaki, R.~Palmer, M.~Zisman, and J.~Gallardo,
  eds.}}}, \bibinfo{type}{Tech. Rep.}, \bibinfo{institution}{{BNL-52623}}
  (\bibinfo{year}{2001}),
  \bibinfo{note}{{\url{http://www.cap.bnl.gov/mumu/studyii/FS2-report.html}}}.

\bibitem[{\citenamefont{Geer}(1998)}]{geer98}
\bibinfo{author}{\bibfnamefont{S.}~\bibnamefont{Geer}}, \bibinfo{journal}{Phys.
  Rev.} \textbf{\bibinfo{volume}{{D57}}}, \bibinfo{pages}{6989}
  (\bibinfo{year}{1998}), \bibinfo{note}{{\textit{ibid.} \textbf{59}, 039903E
  (1999)}}.

\bibitem[{\citenamefont{{M.~M.~Alsharo{\'{}}a~\textit{et
  al.}}}(2003)}]{status_report}
\bibinfo{author}{\bibnamefont{{M.~M.~Alsharo{\'{}}a~\textit{et al.}}}},
  \bibinfo{journal}{Phys. Rev. ST Accel. Beams} \textbf{\bibinfo{volume}{6}},
  \bibinfo{pages}{081001} (\bibinfo{year}{2003}).

\bibitem[{\citenamefont{{A.~Blondel \textit{et al.}}}()}]{blondel}
\bibinfo{author}{\bibnamefont{{A.~Blondel \textit{et al.}}}},
  \bibinfo{howpublished}{{CERN 2004-002 ECFA/CERN}}.

\bibitem[{aps(2004)}]{aps-study}
\emph{\bibinfo{title}{{APS Multi-Divisional Study of the Physics of
  Neutrinos}}},
  \bibinfo{howpublished}{\url{http://www.interactions.org/neutrinostudy/},
  sponsored by the American Physical Society Divisions of: Nuclear~Physics,
  Particles and Fields, Astrophysics, Physics of Beams} (\bibinfo{year}{2004}).

\bibitem[{MC()}]{MC}
\bibinfo{note}{{The Neutrino Factory and Muon Collider Collaboration WEB page,
  \url{http://www.cap.bnl.gov/mumu/}}}.

\bibitem[{UK()}]{UK}
\bibinfo{note}{{\url{http://hepunx.rl.ac.uk/neutrino-factory/}}}.

\bibitem[{CER()}]{CERN}
\bibinfo{note}{{\url{http://muonstoragerings.web.cern.ch/muonstoragerings/}}}.

\bibitem[{JAP()}]{JAPAN}
\bibinfo{note}{{\url{http://www-prism.kek.jp/nufactj/index.html}}}.

\bibitem[{\citenamefont{{N.~Holtkamp and D.~Finley, eds.}}(2000)}]{fs1}
\bibinfo{author}{\bibnamefont{{N.~Holtkamp and D.~Finley, eds.}}},
  \bibinfo{type}{Tech. Rep.} \bibinfo{number}{Fermilab-Pub-00/108-E},
  \bibinfo{institution}{{Fermilab}} (\bibinfo{year}{2000}),
  \bibinfo{note}{{\url{http://www.fnal.gov/projects/muon_collider/nu-factory/n%
u-factory.html}}}.

\bibitem[{\citenamefont{Neuffer}(2003{\natexlab{a}})}]{adiab1}
\bibinfo{author}{\bibfnamefont{D.}~\bibnamefont{Neuffer}},
  \emph{\bibinfo{title}{Exploration of the high-frequency buncher concept}},
  \bibinfo{howpublished}{{MUC-NOTE-269}} (\bibinfo{year}{2003}{\natexlab{a}}),
  \bibinfo{note}{{also, D.~Neuffer, \emph{High-Frequency Buncher and Phase
  Rotation for the Muon Source}, MUC-NOTE-181 (2000). All MUC-NOTE papers are
  available from \url{http://www-mucool.fnal.gov/notes/noteSelMin.html}}}.

\bibitem[{\citenamefont{Neuffer}(2003{\natexlab{b}})}]{adiab2}
\bibinfo{author}{\bibfnamefont{D.}~\bibnamefont{Neuffer}},
  \emph{\bibinfo{title}{Beam dynamics problems of the muon collaboration:
  $\nu$-factory and $\mu^+ - \mu^-$ colliders}},
  \bibinfo{howpublished}{MUC-NOTE-266} (\bibinfo{year}{2003}{\natexlab{b}}).

\bibitem[{\citenamefont{{D.~Neuffer and A.~Van Ginneken}}(2001)}]{adiab4}
\bibinfo{author}{\bibnamefont{{D.~Neuffer and A.~Van Ginneken}}},
  \bibinfo{howpublished}{{Proceedings of the 2001 Particle Accelerator
  Conference}} (\bibinfo{year}{2001}),
  \bibinfo{note}{{\url{http://accelconf.web.cern.ch/Accel/Conf/p01/PAPERS/TPPH%
162.pdf}}}.

\bibitem[{\citenamefont{{A.~Van Gineeken}}(2001)}]{adiab5}
\bibinfo{author}{\bibnamefont{{A.~Van Gineeken}}}, \bibinfo{type}{Tech. Rep.},
  \bibinfo{institution}{Fermilab} (\bibinfo{year}{2001}),
  \bibinfo{note}{{MUC-NOTE-220}}.

\bibitem[{\citenamefont{Fernow}(1999)}]{icool}
\bibinfo{author}{\bibfnamefont{R.}~\bibnamefont{Fernow}}, in
  \emph{\bibinfo{booktitle}{{Proceedings of the 1999 Particle Accelerator
  Conference}}}, edited by \bibinfo{editor}{\bibnamefont{{A.~Luccio and
  W.~MacKay}}} (\bibinfo{year}{1999}), p. \bibinfo{pages}{3020},
  \bibinfo{note}{{latest version available at
  \url{http://pubweb.bnl.gov/people/fernow/icool/readme.html}.}}

\bibitem[{\citenamefont{Mokhov}(2001)}]{mars1}
\bibinfo{author}{\bibfnamefont{N.}~\bibnamefont{Mokhov}}, in
  \emph{\bibinfo{booktitle}{{Proceedings of the 2001 Particle Accelerator
  Conference}}} (\bibinfo{year}{2001}), p. \bibinfo{pages}{745},
  \bibinfo{note}{see also \url{http://www-ap.fnal.gov/MARS/}}.

\bibitem[{\citenamefont{McDonald}(2001)}]{target}
\bibinfo{author}{\bibfnamefont{K.}~\bibnamefont{McDonald}}, in
  \emph{\bibinfo{booktitle}{Proceedings of the 2001 Particle Accelerator
  Conference}} (\bibinfo{year}{2001}), p. \bibinfo{pages}{1583},
  \bibinfo{note}{{also, H.G. Kirk \textit{et al.}, ibid. p. 1535 and Chapter 3
  in ~\cite{fs2}. All Particle Accelerator Conference papers can be obtained
  from \url{http://accelconf.web.cern.ch/accelconf/}}}.

\bibitem[{\citenamefont{{D.~Neuffer}}(2004)}]{NeufferRef}
\bibinfo{author}{\bibnamefont{{D.~Neuffer}}} (\bibinfo{year}{2004}),
  \bibinfo{note}{{presentation at APS Study Workshop, ANL;
  \url{http://www.cap.bnl.gov/mumu/study2a/notes/neuffer.pdf}}}.

\bibitem[{\citenamefont{Chao and Tigner}(1999)}]{multipac}
\bibinfo{editor}{\bibfnamefont{A.}~\bibnamefont{Chao}} \bibnamefont{and}
  \bibinfo{editor}{\bibfnamefont{M.}~\bibnamefont{Tigner}}, eds.,
  \emph{\bibinfo{title}{Handbook of Accelerator Physics and Engineering}}
  (\bibinfo{publisher}{World Scientific}, \bibinfo{year}{1999}).

\bibitem[{\citenamefont{Fernow}(2005)}]{heating}
\bibinfo{author}{\bibfnamefont{R.}~\bibnamefont{Fernow}},
  \emph{\bibinfo{title}{Heating in the study 2a absorber window}},
  \bibinfo{howpublished}{{MUC-NOTE-317}} (\bibinfo{year}{2005}).

\bibitem[{\citenamefont{Symon et~al.}(1956)\citenamefont{Symon, Kerst, Jones,
  Laslett, and Terwilliger}}]{pr103:1837}
\bibinfo{author}{\bibfnamefont{K.~R.} \bibnamefont{Symon}},
  \bibinfo{author}{\bibfnamefont{D.~W.} \bibnamefont{Kerst}},
  \bibinfo{author}{\bibfnamefont{L.~W.} \bibnamefont{Jones}},
  \bibinfo{author}{\bibfnamefont{L.~J.} \bibnamefont{Laslett}},
  \bibnamefont{and} \bibinfo{author}{\bibfnamefont{K.~M.}
  \bibnamefont{Terwilliger}}, \bibinfo{journal}{Phys.\ Rev.}
  \textbf{\bibinfo{volume}{103}}, \bibinfo{pages}{1837} (\bibinfo{year}{1956}).

\bibitem[{\citenamefont{Johnstone et~al.}(1999)\citenamefont{Johnstone, Wan,
  and Garren}}]{pac99:3068}
\bibinfo{author}{\bibfnamefont{C.}~\bibnamefont{Johnstone}},
  \bibinfo{author}{\bibfnamefont{W.}~\bibnamefont{Wan}}, \bibnamefont{and}
  \bibinfo{author}{\bibfnamefont{A.}~\bibnamefont{Garren}}, in
  \emph{\bibinfo{booktitle}{Proceedings of the 1999 Particle Accelerator
  Conference}}, edited by
  \bibinfo{editor}{\bibfnamefont{A.}~\bibnamefont{Luccio}} \bibnamefont{and}
  \bibinfo{editor}{\bibfnamefont{W.}~\bibnamefont{MacKay}}
  (\bibinfo{publisher}{IEEE}, \bibinfo{address}{Piscataway, NJ},
  \bibinfo{year}{1999}), p. \bibinfo{pages}{3068}.

\bibitem[{\citenamefont{Mills and Johnstone}(1999)}]{mumu4:693}
\bibinfo{author}{\bibfnamefont{F.~E.} \bibnamefont{Mills}} \bibnamefont{and}
  \bibinfo{author}{\bibfnamefont{C.}~\bibnamefont{Johnstone}},
  \bibinfo{howpublished}{in the transparency book for the 4$^\text{th}$
  International Conference on Physics Potential \& Developent of $\mu^+$
  $\mu^-$ colliders, San Francisco, CA} (\bibinfo{year}{1999}),
  \bibinfo{note}{{UCLA}, Los Angeles, CA, pp.~693--698}.

\bibitem[{\citenamefont{{M.~Ono \textit{et al.}}}(1999)}]{Ono99}
\bibinfo{author}{\bibnamefont{{M.~Ono \textit{et al.}}}},
  \emph{\bibinfo{title}{{Magnetic field effects on superconducting cavity}}},
  \bibinfo{howpublished}{{9$^{\text{th}}$ Workshop on RF Superconductivity}}
  (\bibinfo{year}{1999}), \bibinfo{note}{{Los Alamos, NM, 2000), Los Alamos
  National Laboratory report, LA-13782-C.}}

\bibitem[{\citenamefont{{J.S.~Berg, C.~Johnstone, and
  D.~Summers}}(2001)}]{pac01:3323}
\bibinfo{author}{\bibnamefont{{J.S.~Berg, C.~Johnstone, and D.~Summers}}}, in
  \emph{\bibinfo{booktitle}{Proceedings of the 2001 Particle Accelerator
  Conference}}, edited by \bibinfo{editor}{\bibnamefont{{P.~Lucas and
  S.~Webber}}} (\bibinfo{year}{2001}), p. \bibinfo{pages}{3323},
  \bibinfo{note}{{D.J.~Summers, Snowmass 2001, \eprint{hep-ex/0208010}}}.

\bibitem[{\citenamefont{Berg}(2004)}]{muc-309}
\bibinfo{author}{\bibfnamefont{J.~S.} \bibnamefont{Berg}},
  \emph{\bibinfo{title}{Recent results from optimization studies of linear
  non-scaling {FFAGs} for muon acceleration}},
  \bibinfo{howpublished}{{MUC-CONF-ACCELERATION-309}, Proceedings of
  {FFAG}\,04, KEK, Tsukuba, Japan, 13--16 October} (\bibinfo{year}{2004}),
  \bibinfo{note}{\url{http://www-mucool.fnal.gov/notes/noteSelMin.html}}.

\bibitem[{\citenamefont{{R.B.~Palmer, J.S.~Berg}}(2004)}]{jsb-cost}
\bibinfo{author}{\bibnamefont{{R.B.~Palmer, J.S.~Berg}}},
  \emph{\bibinfo{title}{{A Model for Determining Dipole, Quadrupole, and
  Combined Function Magnet Costs}}}, \bibinfo{howpublished}{{Proceedings of
  EPAC}} (\bibinfo{year}{2004}).

\bibitem[{ope()}]{opera3d}
\emph{\bibinfo{title}{{Vector Fields Inc., computer program OPERA-3d}}}.

\bibitem[{\citenamefont{{S.~Koscielniak and
  C.~Johnstone}}(2003)}]{KoscielniakPAC03}
\bibinfo{author}{\bibnamefont{{S.~Koscielniak and C.~Johnstone}}}, in
  \cite{pac03}, pp. \bibinfo{pages}{1831--1833}.

\bibitem[{\citenamefont{{S.~Caspi and R.~Hafalia}}(2004)}]{CaspiRef}
\bibinfo{author}{\bibnamefont{{S.~Caspi and R.~Hafalia}}},
  \emph{\bibinfo{title}{{A Combined Function Superconducting Magnet for
  Fixed-Field Muon Acceleration in an Alternating Gradient Ring: First-Cut}}},
  \bibinfo{howpublished}{{LBNL Report SC-MAG-839}} (\bibinfo{year}{2004}).

\bibitem[{\citenamefont{{T.~Nakamoto \textit{et al.}}}({2004})}]{JPARC-magnet1}
\bibinfo{author}{\bibnamefont{{T.~Nakamoto \textit{et al.}}}},
  \bibinfo{journal}{{IEEE Trans. Appl. Supercond.}}
  \textbf{\bibinfo{volume}{{14}}}, \bibinfo{pages}{{616}}
  (\bibinfo{year}{{2004}}).

\bibitem[{\citenamefont{{T.~Ogitsu \textit{et al.}}}({2004})}]{JPARC-magnet2}
\bibinfo{author}{\bibnamefont{{T.~Ogitsu \textit{et al.}}}},
  \bibinfo{journal}{{IEEE Trans. Appl. Supercond.}}
  \textbf{\bibinfo{volume}{{14}}}, \bibinfo{pages}{{604}}
  (\bibinfo{year}{{2004}}).

\bibitem[{\citenamefont{{T.~Nakamoto \textit{et al.}}}({2005})}]{JPARC-magnet3}
\bibinfo{author}{\bibnamefont{{T.~Nakamoto \textit{et al.}}}},
  \emph{\bibinfo{title}{{Development of Superconducting Combined Function
  Magnets for the Proton Transport Line for the J-PARC Neutrino Experiment}}},
  \bibinfo{howpublished}{{to appear in the Proceedings of the 2005 Particle
  Accelerator Conference}} (\bibinfo{year}{{2005}}).

\bibitem[{\citenamefont{{N.~Autin, ed.}}(2000)}]{nufact99}
\bibinfo{author}{\bibnamefont{{N.~Autin, ed.}}}, \bibinfo{journal}{{Nucl.
  Instrum. \& Meth.}} \textbf{\bibinfo{volume}{{A451}}} (\bibinfo{year}{2000}),
  \bibinfo{note}{{Proceedings of the ICFA/ECFA Workshop NUFACT'99: Neutrino
  Factories based on Muon Storage Rings, Lyon France}}.

\bibitem[{\citenamefont{{J.~Norem, \textit{et al.}}}(2003)}]{DarkCurrentnote}
\bibinfo{author}{\bibnamefont{{J.~Norem, \textit{et al.}}}},
  \bibinfo{journal}{{Phys. Rev. ST Accel. Beams}} \textbf{\bibinfo{volume}{6}},
  \bibinfo{pages}{072001} (\bibinfo{year}{2003}), \bibinfo{note}{{also
  MUC-NOTE-226 (2001)}}.

\bibitem[{\citenamefont{Geng et~al.}(2003)}]{newscottref}
\bibinfo{author}{\bibfnamefont{R.~L.} \bibnamefont{Geng}} \bibnamefont{et~al.},
  in  \cite{pac03}, pp. \bibinfo{pages}{1309--1311}.

\bibitem[{\citenamefont{Palmer and Zisman}(2005)}]{MC-322}
\bibinfo{author}{\bibfnamefont{R.}~\bibnamefont{Palmer}} \bibnamefont{and}
  \bibinfo{author}{\bibfnamefont{M.}~\bibnamefont{Zisman}},
  \emph{\bibinfo{title}{{Estimation of Study~IIB costs based on Study~II}}},
  \bibinfo{howpublished}{{MUC-NOTE-322}} (\bibinfo{year}{2005}).

\bibitem[{\citenamefont{{M. A.~Green, R.~Byrns,
  S.J.St.~Lorant}}(1992)}]{MAGref}
\bibinfo{author}{\bibnamefont{{M. A.~Green, R.~Byrns, S.J.St.~Lorant}}},
  \bibinfo{journal}{{Advances in Cryo. Eng.}} \textbf{\bibinfo{volume}{37}},
  \bibinfo{pages}{637} (\bibinfo{year}{1992}), \bibinfo{note}{{LBNL-30824; see
  also Phys. Rev. \textbf{D66}, Review of Particle Physics, 010001-217
  (2002)}}.

\bibitem[{\citenamefont{{J. S.~Berg, R.~Fernow, and R.
  B.~Palmer}}(2004)}]{InjExtRef}
\bibinfo{author}{\bibnamefont{{J. S.~Berg, R.~Fernow, and R. B.~Palmer}}},
  \bibinfo{howpublished}{{FFAG Workshop, Vancouver, Canada}}
  (\bibinfo{year}{2004}), \bibinfo{note}{{see
  \url{http://www.triumf.ca/ffag2004/.}}}

\bibitem[{\citenamefont{Chew et~al.}(2003)\citenamefont{Chew, Lucas, and
  Webber}}]{pac03}
\bibinfo{editor}{\bibfnamefont{J.}~\bibnamefont{Chew}},
  \bibinfo{editor}{\bibfnamefont{P.}~\bibnamefont{Lucas}}, \bibnamefont{and}
  \bibinfo{editor}{\bibfnamefont{S.}~\bibnamefont{Webber}}, eds.,
  \emph{\bibinfo{title}{Proceedings of the 2003 Particle Accelerator
  Conference}} (\bibinfo{publisher}{IEEE}, \bibinfo{address}{Piscataway, NJ},
  \bibinfo{year}{2003}).

\end{thebibliography}
\end{document}